\pdfoutput=1

\documentclass[acmlarge]{acmart}

\makeatletter
\newcommand{\myconfshort}{\acmConference@shortname}
\newcommand{\myconffull}{\acmConference@name}
\graphicspath{{images/}{emo-graphics/}}
\newcommand{\myconfdate}{\acmConference@date}
\newcommand{\myconfloc}{\acmConference@venue}
\AtBeginDocument{
  \fancypagestyle{firstpagestyle}{
    \fancyhead{}\fancyfoot[C]{}}
  \fancyhf{}
  \fancyhead[LO]{\@headfootfont\shorttitle}\fancyhead[RE]{\@headfootfont\@shortauthors}\fancyhead[LE]{\@headfootfont\footnotesize \myconfshort, \myconfdate, \myconfloc}\fancyhead[RO]{\@headfootfont\footnotesize \myconfshort, \myconfdate, \myconfloc}\fancyfoot[C]{}}
\makeatother
\acmBooktitle{\conffull\@ (\confshort), \confdate, \confloc}

\usepackage{epigraph}
\usepackage[extra]{emo}

\usepackage{iftex}
\ifLuaTeX
\usepackage{emoji}
\else
\newcommand\emoji[1]{\emo{#1}}
\fi
\usepackage{enumitem}
\usepackage{xcolor}
\usepackage{graphicx}
\usepackage{float}
\copyrightyear{2026}
\acmYear{2026}
\setcopyright{cc}
\setcctype{by-nc-nd}
\acmConference[FAccT '26]{The 2026 ACM Conference on Fairness, Accountability, and Transparency}{June 25--28, 2026}{Montreal, QC, Canada}
\acmBooktitle{The 2026 ACM Conference on Fairness, Accountability, and Transparency (FAccT '26), June 25--28, 2026, Montreal, QC, Canada}
\acmDOI{10.1145/3805689.3812392}
\acmISBN{979-8-4007-2596-8/2026/06}

\AtBeginDocument{\libertineOsF}

\newlist{EnumerateLining}{enumerate}{1}
\setlist[EnumerateLining]{label=\libertineLF(\arabic*)}
\newcommand\REF[1]{\S\kern0.07em\ref{#1}}

\begin{document}

\newcommand\V[1]{\textsc{\MakeLowercase{#1}}}

\newcommand\DDAALLEE{\texorpdfstring{DALL\raisebox{0.2ex}{•}E}{DALL•E}}  \newcommand\DALLE{\textsc{dall•e}}

\title{Mapping the Stochastic Penal Colony \emoji{desert-island}} 

\author{Robert Grimm}
\orcid{0000-0002-8300-2153}
\affiliation{\institution{Charles University}
    \city{Prague}
    \country{Czech Republic}
}
\email{rgrimm@alum.mit.edu}



\begin{abstract}
With peak content moderation seemingly behind us, this paper revisits its
punitive side. But instead of focusing on who is being (disproportionately)
moderated, it focuses on the punishment itself and explores the question of how
content moderation treats users posting violative content unjustly, while the
organizations doing the moderation act in a self-serving manner. First, this
paper reworks Foucault's model of the penal system for the algorithmic age,
restoring the penal colony as a figuratively liminal practice between punishment
as performance and punishment as discipline, i.e., the stochastic penal colony.
Second, it develops a novel methodology that combines auto-ethnography for
collecting experiences and artifacts with procedural justice for analyzing them.
Third, it applies this conceptual and methodological framing to three case
studies, one on pre-Musk Twitter's gallingly performative moderation, one on
OpenAI's exhaustively controlling moderation for \DALLE~2, and one on
Pinterest's underhandedly manipulative moderation. While substantially
different, all three feature the pervasive threat of account suspension, which
banishes users to the stochastic penal colony.
 \end{abstract}

\maketitle

\newenvironment{kafkaesque}{
    \setlength{\parindent}{\normalparindent}
}{}
\renewcommand{\textflush}{kafkaesque}
\setlength{\epigraphwidth}{0.8\textwidth}

\epigraph{As you see, it consists of three parts. With the passage of time
certain popular names have been developed for each of these parts. The one
underneath is called the Bed, the upper one is called the Inscriber, and here in
the middle, this moving part is called the Harrow. [\ldots]

As soon as the man is strapped in securely, the Bed is set in motion. It quivers
with tiny, very rapid oscillations from side to side and up and down
simultaneously. [\ldots] Only with our Bed all movements are precisely
calibrated, for they must be meticulously coordinated with the movements of the
Harrow. But it's the Harrow which has the job of actually carrying out the
sentence. [\ldots]

The law which a condemned man has violated is inscribed on his body with the
Harrow.}{Franz Kafka, \emph{In the Penal Colony}~\cite{Kafka1995}}

\section{Introduction}
\label{sec:introduction}

Folk theories---``Facebook jail''---as well as some legal
scholars~\cite{Klonick2018} rather optimistically imagine content moderation and
similar algorithmic processes, including resulting punishments, as a rational
and orderly governance function. Yet, both academic and grey literature track an
abundance of often discriminatory and punitive algorithmic interventions by
corporations and universities~\cite{Bucher2018,Eubanks2018,NarayananKapoor2024}.
Examples include credit scoring~\cite{Anonymous2018}, debt
assessment~\cite{Yampolskiy2015}, exam proctoring~\cite{FrancisWard2021b}, fraud
prevention~\cite{Kugel2022}, grading~\cite{Lam2020}, job and school
applications~\cite{Anonymous2016,Hall2012,Hall2020a,Stockton2020}, personal
vendettas~\cite{Casovan2022}, private security services~\cite{HaoSwart2022},
productivity monitoring~\cite{Covert2022, HaoFreischlad2022,
KantorSundaramea2022, Rosenblat2018}, and screening for child sexual abuse
materials~\cite{Atherton2022a}. Still, Amazon's warehouses stand out for their
profit-driven amorality~\cite{KantorWeiseea2021,Lennard2020}: The firm's
algorithmic exploitation of its workers not only leads to high injury
rates~\cite{Brown2019a,Clark2023,Sainato2021}, but the attendant 150\% yearly
staff turnover means that Amazon may run out of people to exploit roughly about
now~\cite{Sainato2022}.

To explore the many ways such algorithmic interventions end up being unjust at
best and punitive at worst, this paper focuses on the algorithmic intervention
most commonly deployed at large, content moderation. In particular, we start by
restoring banishment to a penal colony as an intermediate practice in Michel
Foucault's structural analysis of punishment~\cite{Foucault1979}, historically
falling between punishment as public spectacle and punishment by disciplinary
institutions. Since the threat of banishment also is ubiquitous in content
moderation, we then propose the \emph{stochastic penal colony} as a conceptual
model for the punitive impact of algorithmic content moderation---albeit one
with clearly lower stakes than the historical penal colony. To make this model
specific, we further draw on procedural justice~\cite{Tyler2003, Tyler2006,
Tyler2007} and define the stochastic penal colony as automated interventions
that routinely inflict injustices on users by stifling their voice, curtailing
their agency, and violating their dignity, while moderating organizations are
far from neutral or trustworthy actors.

We rely on auto-ethnography~\cite{Ellis2003} to surface templated and ephemeral
materials for the punishment processes of pre-Musk Twitter in October 2021,
OpenAI's \DALLE\ in July 2022, and Pinterest in March 2024. After combining them
with relevant policy documents, we then use the above definition of the
stochastic penal colony for analyzing the three corporations' content
moderation. The three case studies present substantially different results for
each of them. Twitter ends up othering and degrading users by staging punishment
as a performance. But because it is limited to individual users only, Twitter
can avoid the potential blowback from a public spectacle. Next, OpenAI's \DALLE\
maximizes control by strictly enforcing an unreasonably expansive policy and
falling back on post-hoc social and legal means to handle violative content that
might have slipped through. Finally, Pinterest manufactures a sense of severity
and urgency through a dedicated web interface for communicating violative
content. Yet, it also seems to struggle with improving its classifier for adult
content and proactively communicating changes in its enforcement. Finally,
common to all three case studies is the pervasive and often ill-defined threat
of banishment.

The contributions of this paper are threefold: First, we extend procedural
justice with a fifth criterion, agency, in addition to the original criteria of
voice, dignity née respect, neutrality, and trust. Agency may not play a
significant role in the original context of procedural justice, sovereign law
and justice. But that changes when considering corporate or academic governance
processes. Second, we explore the conceptual, political, historical, and
literary context of the penal colony and position it as a suitable liminal model
for overly punitive interventions. We also provide a concrete definition based
on the five criteria of procedural justice. Third, we use this definition across
three case studies on content moderation by Twitter, OpenAI's \DALLE, and
Pinterest and demonstrate that, while they do not go as far as turning the
elderly into mulch~\cite{KeyesHutsonea2019}, all three commit major injustices.
At the same time, they differ significantly in their overall emphasis as well as
particulars.
 \section{The Penal Colony: Conceptual, Political, Historical, and Literary Context}
\label{sec:context}

In \emph{Discipline and Punish}, Michel Foucault traces the transition from
punishment as a public and usually deadly spectacle to the modern prison and
other disciplinary institutions~\cite{Foucault1979}. He argues that this
transition did not happen for humanist concerns, as the result of reform
efforts. Instead, the driving force was the destabilizing impact of public
executions. By being rather ostentatious displays of power, they turned the
criminal into sympathetic victim. In contrast, executioner as well as sovereign
became targets of popular resentment. By simultaneously rationalizing,
tempering, and distributing the application of power, penal institutions avoid
these downsides. They instead instill discipline into the individual under their
custody. As people internalize discipline, that self-discipline obviates the
need for more direct applications of power and begets other disciplinary
institutions, including schools, hospitals, and factories.

The institutionalization of discipline does place constraints on the sovereign's
exercise of power and leads to an attendant loss of centralized control.
Notably, Foucault's idealized disciplinary institution, the panopticon, i.e., a
circular arrangement of cells around a central monitoring station or tower,
simply can't scale beyond maybe 500 cells---at least in a domain where gravity,
mass, stress, and strain reign supreme. Arguably, East Germany was a notable
exception. But its lo-tech approach to central control also was too expensive to
be sustainable~\cite{Schroeder2013}. For instance, in one district, 18\% of the
population were active informants for the Stasi, that country's vicious state
security service~\cite{Kellerhoff2022}.

Here too, computing technology is proving to be a game changer---for the
worse: China under its current, particularly authoritarian president Xi Jinping,
is reaping the benefits of readily available hardware sensors and machine
learning algorithms. By rolling out ever more intrusive yet centralized control,
China is erasing the distinction between prison and not-prison at
scale~\cite{Grauer2021,MozurXiaoea2022,SmithIV2016}. Ironically, some of that is
driven by American innovations on predictive
policing~\cite{PerryMcInnisea2013,SmithIV2016,Sprick2019}. But the excessive
intrusiveness also makes China's \emph{surveillance state} an unsuitable model
for algorithmic control in Western democracies---for now.

The \emph{carceral state} in the United States comes somewhat
closer~\cite{Simon2007}. After all, it is an early and aggressive adopter of
algorithmic enforcement technologies~\cite{AngwinLarsonea2016, EPIC2020,
Hao2019, ReddenODonovanDixea2020, Yampolskiy2016}. But its relentless focus on
incarceration---in 2022, the country accounted for 20\% of imprisoned people
across the world~\cite{SawyerWagner2022} but only 4.25\% of the world
population~\cite{Worldometer2023}---also makes it an extreme and unsuitable
model for corporations and other civil institutions, which, as listed in
\REF{sec:introduction}, deploy overly punitive but non-carceral
interventions.

In doing so, these institutions may even innovate on punishment. As this paper's
case studies will demonstrate, Twitter's and Pinterest's enforcement processes
revive aspects of punishment as performance, but do so while (ingeniously)
avoiding the destabilizing public spectacle. Meanwhile, OpenAI's \DALLE\
combines preemptive algorithmic enforcement with post-hoc offline mechanisms to
maximize control. Meanwhile, the threat of banishment as punishment is
omnipresent across all three case studies. This focus on the punitive potential
outside the control of the sovereign state also distinguishes this paper from
related work exploring the same in the context of the surveillance and carceral
states~\cite{DehlendorfGerety2021,McElroyWhittakerea2021}.

Because of the prominent role of banishment across the case studies, we are
proposing the \emph{penal colony} as a fitting model for contemporary
algorithmic practices outside the criminal justice system. The French version of
\emph{transportation}---the practice of sending prisoners to far off
locales---is far more recent than we'd probably like to acknowledge. France
began turning French Guiana into one large penal colony from 1852
onwards---after the British had already begun unwinding their own penal
colonies---and closed the colony only in
1953~\cite{Aldrich2010,Anderson2018,Spierenburg2009}. For the 70 or so years
before closing, transportation was reserved for convicts sentenced under
France's own three strikes laws. It also was almost always terminal: Only 2,000
out of 70,000 prisoners returned to France during their
lifetimes~\cite{WallechinskyWallace1978}. For that reason, prisoners referred to
the penal colony as ``dry guillotine''~\cite{Furlong1913,ReneBelbenoit1938}. Yet
discipline was inconsistent, even lax, depending on location.

Foucault had surprisingly little to say about the penal
colony~\cite{Redfield2005}, even though transportation must be understood as a
distinct intermediate, or liminal, stage in penal history. As such, it combines
aspects from the performance of punishment and the discipline of prisons.
Notably, like earlier practices, transportation is usually terminal. But unlike
earlier practices, the penal colony is conveniently out of sight. The penal
colony also incorporates a disciplinary component, typically involving hands-on
labor to create the infrastructure for more general colonization. (Against that
background, it seems rather fitting that French Guiana nowadays serves as the
Europen Union's launching pad for the colonization of space.)

While the concept of the \emph{stochastic penal colony} is inspired by the
historical penal colony, there also are pronounced qualitative differences. Most
importantly, the stakes are much lower for the stochastic variant. Banishment in
the stochastic version may limit access to platforms or resources, whereas
banishment in the historical version was a matter of life and death.
Furthermore, the stochastic version necessarily favors punitive interventions
over physical discipline or forced labor. Next, the stochastic penal colony
utilizes some of the same \V{AI} technology as China's surveillance state and
the United States' carceral state. Yet, there is no central control, or even
intent. Its downsides, not surprisingly, also worsen along, for example, class
and racial lines, but it ensnares the privileged, including white people, almost
as easily.

Despite the lower stakes, banishment into the stochastic penal colony can still
have (negative) life-altering impact, for example, when people depend on social
media for their livelihood as influencers, solo practitioners, or small business
owners. Furthermore, given economic concentration in the technology
sector---with, for example, Meta operating five social media platforms and Match
Group operating over 40~dating services---banishment from one platform may
really be banishment from many platforms and may leave affected individuals with
few, if any, other options for equivalent services. The downsides tend to get
only more dire for most of the other applications of algorithmic control listed
in~\REF{sec:introduction}.

In contrast to sovereign justice in Western democracies, for which
constitutional provisions guarantee \emph{human judgement}, the stochastic penal
colony is almost entirely mechanized. Machine learning models serve as judge and
jury, while scripted processes currently serve as executioner. Even appeals are
mostly administered by automated, \V{AI}-based processes. The overall result is
fundamentally \emph{inhuman governance}, which arguably also ensures that
punishment becomes inhumane.

Even though Foucault largely ignored the penal colony in his analysis of
discipline and punishment, his basic argument applies to the stochastic penal
colony just as well. It too was not introduced for humanist concerns, but to
impose additional discipline, further narrowing what is permissible speech and
conduct online. It also obscures the accumulation of resources and power by a
few technology firms. For instance, in 2025, seventh-ranked Meta had higher
revenues than the \V{GDP} of all but 60 countries, third-ranked Alphabet had
higher revenues than the \V{GDP} of all but 44 countries, and top-ranked Amazon
had higher revenues than the \V{GDP} of all but 26
countries~\cite{WikipediaTechFirmsByRevenue2025, WikipediaCountriesByGDP2026}.
Meanwhile, technology corporations hawk their algorithmic prowess for
``conjuring'' results~\cite{NagyNeff2024}, which further discourages us humans
from questioning, let alone doubting, their quality and accuracy.

The remoteness of the penal colony, both literally and figuratively, also turns
it into an effective, intellectual investigative device that renders
contemporary practice strange again and hence amenable to analysis. While this
conceptual take on the penal colony is largely ahistorical, it nonetheless
traces right back to French Guiana. As indicated by this paper's opening quote,
the stochastic penal colony is also inspired by Franz Kafka's 1919 short story
\emph{In the Penal Colony}~\cite{Kafka1995}. Kafka, in turn, was
influenced~\cite{Robertson2017} by Octave Mirbeau's 1899 novel \emph{The Torture
Garden}~\cite{Mirbeau2008}. While taking place in an imaginary China, its year
of publication and dedication---\emph{To Priests, Soldiers, Judges / to men who
rear, lead, or govern men / I dedicate these pages of murder and blood}.---point to the Dreyfus affair as primary inspiration. Alfred Dreyfus, a Jew and
French military officer, had been falsely convicted for espionage in early 1895
and again in 1899---with rampant antisemitism leading to the systematic
suppression of exculpatory evidence and complete disregard of the real spy's
public confession in 1898. As a result, Mr Dreyfus spent 1895--1899 on Devil's
Island, a particularly harsh site in the French-Guianan penal colony.
Coincidentally, the Dreyfus affair also popularized the word ``intellectual,''
albeit starting out as a pejorative~\cite{Drake2005,
StudentsAtTheUniversityOfBristol2021}.

Besides, the stochastic penal colony~\emoji{desert-island} provides an excellent
habitat for a pandemonium of stochastic
parrots~\emoji{parrot}~\cite{BenderGebruea2021}!
 \newpage
\section{Methods}
\label{sec:methods}

This paper explores the following research question:

\begin{quote}
\textit{In what ways do Twitter, OpenAI's \DALLE, and Pinterest treat users
posting violative content unjustly, while also acting in a self-serving manner?}
\end{quote}

\noindent{}Studying these punishment processes requires:

\begin{EnumerateLining}
\item Surfacing templated and ephemeral communications, which are only available
to targeted users, and then combining them with publicly available corporate
policies into complete workflows.
\item Analyzing the thusly restored complete record of punishment processes
according to a set of well-defined criteria.
\end{EnumerateLining}

To address the two methodological requirements, the three case studies draw on
auto-ethnography and procedural justice, respectively. Auto-ethnography---``to
describe and systematically analyse personal experience in order to understand
cultural experience''~\cite{Ellis2003}--is often employed for exploring marginal
perspectives, oral traditions, and otherwise sensitive practices. But it also
poses significant challenges~\cite{Edwards2021, Tolich2010} to objectivity,
privacy, and consent---especially when a study starts organically instead of
following a formal research protocol or seeks retroactive consent.

For the three case studies, the use of auto-ethnography is largely limited to
observing and documenting mechanized and hence standardized corporate punishment
processes, including templated and ephemeral communications as well as relevant
platform policies. Artifacts were collected by the author as he was subjected to
the punishment processes. They are documented in
appendices~\ref{app:twitter},~\ref{app:dalle}, and~\ref{app:pinterest},
respectively.

Other than violating the author's privacy by outing him as a some time ``daily
active shithead''~\cite{Sherman2021}, this paper's use of auto-ethnography
avoids the above mentioned pitfalls. First, by observing mechanized content
moderation workflows with very few steps and branches, the three case studies
\emph{are} knowledge-generating. Surfacing more complex workflows would require
switching to crowd-sourcing, which effectively becomes the plural of
auto-ethnography. Second, the violative content for the case studies stands on
its own, i.e., is not part of a larger conversation, and only mentions other
people by their professional roles, i.e., oil company \V{CEO}s for the Twitter
case study and the Pope for the \DALLE\ case study. That obviates ethical
approval. A query to the review board at the author's previous institution
confirmed as much. Still, the three case studies are fundamentally based on the
author's experiences. For academic integrity, we thus dispense with the
\textit{pluralis investigatoris} for their write-up and instead fall back on the
auto-ethnographic \textit{I}, including for the analysis.

Carolina Are has similarly employed auto-ethnography for exploring outcomes from
content moderation~\cite{Are2023}. However, compared to her investigation of
Instagram's and TikTok's (lack of) platform affordances, this paper more
carefully limits the use of auto-ethnography to dissecting the punishment
process itself. Otherwise, the two papers are largely complimentary.

To make sense of surfaced processes and materials, the three case studies rely
on procedural justice~\cite{Tyler2003,Tyler2006,Tyler2007}. Notably, we label
violations of a user's voice, agency, and dignity as injustices. After
consulting additional source material, we also account for violations of
platform neutrality and trust as injustices. By comparison, Tyler's original
definition of procedural justice positions voice and neutrality for evaluating
the process of justice, while respect and trust are positioned for evaluating
the relationships between participants. In other words, this paper's application
of procedural justice replaces ``respect'' with ``dignity'' and newly introduces
``agency'' in addition to voice.

Voice means being listened to and given due consideration. Agency means the
freedom to make one's own decisions and then act on them, with an emphasis on
the acting part. In other words, agency implies initiative whereas voice does
not. The omission of agency from the original definition of procedural justice
isn't too surprising given its original context, namely sovereign law and
justice. After all, in the governmental application of (criminal) justice,
individuals' agency is at best a secondary concern and, for the most part,
lacking. But in the context of corporate governance of user-generated content,
the positive presence of agency becomes an important concern.

A more subtle difference is the grouping of the five criteria. Whereas the
original definition of procedural justice distinguishes between process and
relational criteria, we distinguish between criteria that stand on their own and
hence are meaningful when applied to just a single governance process, i.e.,
voice, agency, and dignity, as well as criteria that require additional context,
i.e., neutrality and trust. Clearly, a negative individual experience will
influence one's evaluation of neutrality and trust as well. But being confident
in statements about neutrality and trust when it comes to governance procedures
fundamentally requires more than just one exemplar.

Having clarified the differences in criteria, we can now give a more precise
definition of the stochastic penal colony: It's the human impact resulting from
algorithmic interventions outside the governmental application of justice that
routinely violate voice, agency, dignity, neutrality, or trust. That in turn
also helps clarify the distinction from Foucault's disciplinary institutions:
The latter imply individual compromises or restrictions when it comes to voice,
agency, and dignity. But in the modern conception, they also require
institutional neutrality and trust. Routine violations of the latter two turn a
disciplinary intervention into an injustice.
 \section{October 2021: Tweet-Da-F\'e}
\label{sec:tweet-da-fe}

Early one morning in October 2021, I had just finished reading an article about
some oil industry association spending millions of dollars on lobbying and
advertising to derail the Biden administration's push for climate change
legislation~\cite{Tabuchi2021}. Additionally, three of the association's larger
member companies spent millions of dollars each towards that same goal---despite also being responsible for 8.7\% of all global CO$_2$ emissions since
1965~\cite{TaylorWatts2019}. I was enraged. To vent, I composed a caustic tweet
that @-mentioned the three firms and stated that I was looking forward to their
\V{CEO}s facing capital punishment for genocide. I was well aware of the
statement's severity and incivility while writing it. But I reassured myself
that this was ok, since the statement implied a formal, legal process that still
is practiced in the United States. (I also included that argument in both of my
appeals.)

I remain ambivalent about the tweet. With ExxonMobil's internal projections from
1977 to 2003 ``accurately forecasting warming that is consistent with subsequent
observations''~\cite{SupranRahmstorf2023}, with birds falling dead from the
skies~\cite{Dave2022}, 11 billion crabs just vanishing~\cite{Olmstead2022}, as
well as a third of Pakistan flooding~\cite{Chughtai2022}, and with projections
another twenty years ahead generally devastating~\cite{VannRNewkirkII2025}, all
because of climate change, it is hard \emph{not} to wish harm on responsible
parties including oil companies and their \V{CEO}s. Yet, the record-setting
execution spree towards the end of Donald Trump's first
presidency~\cite{Arnsdorf2020, Kovarsky2022, SuebsaengReis2023} suggests that
tweet and oil company greed are grounded in the same basic inhumanity. Hence,
the screenshot in \REF{app:twitter-staging} obscures the exact text of the
tweet.

The tweet's incivility certainly triggered Twitter's \V{AI}. Within a couple of
seconds after posting, it removed the tweet and locked my account. The stated
justification was pretty specific:

\begin{quote}
\textbf{Violating our rules against
\href{https://web.archive.org/web/20220905021323/https://help.twitter.com/en/rules-and-policies/abusive-behavior}{abuse
and harassment}.}

You may not engage in the targeted harassment of someone, or incite other people
to do so. This includes wishing or hoping that someone experiences physical
harm.
\end{quote}

\noindent{}The linked policy on abusive behavior, reproduced in
\REF{app:twitter-abusive-behavior}, is not only specific but genuinely
helpful. It is written in accessible, well-structured prose: The policy starts
with a rationale, is followed by the different kinds of abusive content, and
concludes with a range of possible sanctions. The mid-section on kinds of
abusive content features well-delineated and reasonable prohibitions. It even
reassures readers that the firm is well aware that some tweets, by themselves,
may appear to violate the policy but, when considered in their original context,
do not.

Thanks to the effective presentation, finding the concrete prohibition
applicable to my tweet was easy: ``Wishing, hoping, or calling for serious harm
on a person or group of people.'' After elaborating on possible context and
giving examples, the policy---rather reasonably---allows that some wishes of
harm may be justified, in the heat of the moment, as expressions of outrage. In
such cases, Twitter still requires offending tweets to be deleted but does not
impose penalties. Apparently, rapists and child abusers count as legitimate
targets but oil company \V{CEO}s do not---yet.

I appealed the decision by Twitter's \V{AI} that same morning. Or at least, I
tried to: Twitter's form for filing an appeal seemed to have the same character
limit as a tweet. That excludes most appeals besides a succinctly stated single
reason. Alas, my justification was far from that and, not surprisingly, Twitter
rejected the appeal three days later. However, the form email notifying me of
the rejection wasn't even filled in, despite containing instructions in \V{HTML}
comments. Since I had located another page for launching an appeal that wasn't
marred by the original form's character limit, I tried again with that form,
this time focusing mostly on the bad form of the rejected appeal. When that
second appeal went unanswered for three weeks, I gave up. I withdrew my appeal,
acknowledged that I ``violated the Twitter Rules,'' and deleted the offending
tweet---all with one click on a big red ``Delete'' button.

Alas, residual effects from the episode remained. When I tried to sign up to
Twitter for Professionals several months later, I got a notification that
``something's missing,'' even though my account met all criteria stated in
Twitter's documentation. Meanwhile, a satirical account of mine, which I opened
more recently and which described my alter ego as a ``lifelong practitioner of
faggotry, promoter of the gay agenda, and unrepentant socialist monarchist,''
could sign up to Twitter for Professionals within days of account creation.
\emph{Très professionnel} indeed!

\subsection{A Punishing Performance}

Alas, Twitter's eminently reasonable policy was only a facade for selecting
candidate users, or \emph{Condemned} in the most general meaning of the word,
for the firm's personalized performance of punishment.

As illustrated in \REF{app:twitter-staging}, the set design was rather
crude: The violative tweet was featured prominently on screen and demarcated the
extent of the Condemned's Twitter for the duration of this performance. While
Twitter's email notifying me of the violation claimed that, ``while in this
state, you can still browse Twitter, but you're limited to only sending Direct
Messages to your followers---no Tweets, Retweets, Fleets, follows, or likes''
(capitalization theirs), that was plainly false. The Condemned's Twitter brooks
no other content or interaction but staring at the offending tweet.

While the set design lacks subtlety, it also is quite effective. It reminds the
audience of the very transgression that started this performance. It also
reminds the audience of the only certain way out of one-tweet-limbo---admitting
the violative character of the tweet and then deleting it. And it reminds the
audience of the final arbiter of account access (or lack thereof): Twitter and
Twitter only. The set design also is rather versatile. By having a well-defined
visual and attentive center, incidental text and \V{UI} widgets surrounding the
one tweet that no one else can see may change without distracting from the
overall message. Hence, after clicking ``cancel your appeal,'' the text below
the one tweet that no one else can see turned into an acknowledgement of guilt
combined with a button to ``delete'' that last vestige of violative content.

In this context, calling that digital artifact a ``tweet'' and having the
Condemned ``delete'' said tweet is largely farcical---also coercive, punitive,
and somewhat degrading. After all, the tweet has long been purged from the
platform by the one entity that has total control over what content is publicly
visible, Twitter. In all likelihood, the tweet's current starring role isn't
harbinger of future virality to come, but rather its last hurrah before
permanent cancellation. The farcical, coercive, punitive, and somewhat degrading
character of the performance makes for four injustices in one. It also makes for
a resounding lack of dignity and respect afforded to the Condemned by
Twitter---which may just explain the surprising emphasis on just those two
qualities exhibited by former Condemned in a survey on procedural justice on
Twitter (see \S\,5.5 in~\cite{KatsarosTylerea2022}).

Twitter's punishing performance features one more twist: The Condemned forms
both the audience and cast of one. The mechanization of content review via
\V{AI} has made the individualized targeting of just one user per performance
cost-effective. The limited outcomes nonetheless enabled the Condemned to make
one substantive choice. They got to determine the duration of the performance:
hours if they forgo appeal, days if they appeal, or forever if they walk out.
Alas, the exact meaning of ``appeal'' in the previous sentence is unclear. Since
Twitter limited justifications to 280 characters, kept admonishing that ``you
won't be able to access your Twitter account'' and to ``just delete your
content,'' provided no explanation for rejecting an appeal, and disclosed no
statistics in its semiannual transparency reports, ``appeal'' became an
unappealing husk of its usual self.

Like Twitter's intervention, the Catholic
Inquisition~\cite{Lea1906a,Lea1906b,Lea1906c,Lea1906d} and Maoist denunciation
rallies~\cite{Yang2021} are centered around a carefully staged assertion of
institutional might and the ritualized subjugation of transgressive individuals.
At the same time, the virtuality of the internet prevents the physical excesses
of these historical precedents, and the scripted personalization avoids their
power-eroding long-term impact, as identified by Foucault. That same
individualized targeting also makes Twitter's punishment process resilient to
outside interference. After all, the vast majority of Twitter users will never
experience such a punishing performance. Furthermore, if pressed, they can
reassure themselves with Twitter's eminently reasonable policies. Yet, any
expression of hurt or anger makes the Condemned come across as clearly
unreasonable or worse---and hence so much easier to dismiss and ignore.

Still, the lack of physical force and torture does raise the question of why
anyone would put up with that shit. The reason was pre-Musk Twitter's rather
unique position as breaking news service, political townsquare, professional
society, and corporate customer service platform in one. Thanks to that
combination, the threat of account termination was substantial and, depending on
a user's Twitter presence, could approach something like real-world social
death. However, thanks to Mr Musk's ``extremely hardcore'' leadership since
taking over the firm~\cite{SchifferNewtonea2023}, Twitter lost plenty of users
and advertisers. Worse, Mr Musk insists not only on running the social network
according to his ever-changing whims, but also must be the most visible user,
dominating notifications. Hence, a return to old form seems unlikely.

\subsection{Twitter's Neutrality and Trustworthiness}
\label{sec:trusting-twitter}

The cognitive dissonance between Twitter's measured policy and its punishing,
performative enforcement may seem extreme at first and make one wonder about the
kind of (dysfunctional) firm culture that tolerates such obviously divergent
practices. But it doesn't take much to get there. The very dehumanizing
condescencion engendered in the punishing performance points to this being just
another case of othering, of us versus them. Twitter employees felt like the
good guys keeping daily active shitheads in check, which licensed them to
dehumanize the shitheads. Nonetheless, the impact of this dissonance is deeply
corrosive and raises significant doubts about Twitter's trustworthiness and
integrity.

It doesn't help that content policies, their enforcement, and their transparency
data are almost entirely silent on a critical salient feature. They hardly
mention the use of \V{AI}. Yet that use is not new and dates back to the
beginning of the pandemic at the very least~\cite{ScottKayali2020}. Clearly, the
firm had plenty of time to update its documentation. Worse, that omission isn't
limited to content policy etc, but extends to \emph{all} of Twitter's help
pages. Table~\ref{table:search} quantifies the number of results from searching
for common variations of the term ``\V{AI}'' using Twitter's own search function
in October 2022. The darth of relevant material a year later is striking. Not
only were there hardly any mentions, but existing ones amounted to little more
than acknowledgements that, for instance, top tweets, topics, and
recommendations were curated algorithmically. There certainly were no
context-providing dataset, model, or system cards to be
found~\cite{GebruMorgensternea2021,MitchellWuea2019,ProcopeCheemaea2022}. That
is inconsistent with Twitter's stated commitment to implementing the Santa Clara
Principles, which require detailed disclosure of automated content
moderation~\cite{AccessNowACLUFoundationOfNorthernCaliforniaea2021}.

\begin{table}
\caption{Search terms and number of hits on Twitter's help pages (as of 21
    October, 2022)}
\label{table:search}
\libertineLF
\begin{tabular}{lr} \toprule
\textbf{Search Term} & \textbf{Results} \\ \midrule
AI & 0 \\
algorithm & 5 \\
artificial intelligence & 1 \\
machine learning & 3 \\ \bottomrule
\end{tabular}
\end{table}

Twitter's transparency report nonetheless helps confirm an important aspect of
its automated content review, namely the exact timing. When my account was
blocked in October 2021, the notification thereof was nearly instantaneous after
posting, but I wasn't entirely sure whether Twitter's application had actually
confirmed the posting of the tweet. This matters since Twitter reviewing all
content before posting also eliminates any notion of human harm and thereby
undermines the justification for any punishment (a point I also raised in my
appeals). Alas, it appears that Twitter's systems performed posting and
reviewing tasks in parallel. In its transparency report for 2021, Twitter used
the rather imprecise buckets of <100, 100--1,000, and >1,000 views before
content removal~\cite{Twitter2022}. In contrast, Pinterest used buckets 0, 1--9,
10--100, and >100 views~\cite{Pinterest2022} and YouTube used buckets 0, 1--10,
and >100 views~\cite{Google2022}. Clearly, the latter two social media were
confident in their proactive content removal, whereas Twitter was not.
 \section{July 2022: \DDAALLEE\ 2 Supermax}
\label{sec:dalle}

My first interaction with \DALLE~2, OpenAI's then ground-breaking text-to-image
system, in late July 2022 didn't quite go as expected. I had signed up for the
service several months before but had been granted access only earlier that day.
So I was eager to try out the system and started with a prompt that had yielded
fascinating results with another text-to-image system:

\begin{quote}
The crucified pope, painting by Francis Bacon
\end{quote}

But instead of producing four new masterworks by the famous
20\textsuperscript{th} century painter that combine two of his most prominent
themes~\cite{Wikipedia2023}, \DALLE~2 responded with a stern warning:

\begin{quote}
It looks like this request may not follow our content policy. Further policy
violations may lead to an automatic suspension of your account.
\end{quote}

\noindent{}As stated in the first sentence, the alleged policy violation is just
that, a possible violation of some policy. The sentence's lack of certainty
(using ``may not'' instead of ``does not'') and specificity (using the generic
``content policy'') were not actionable---beyond not submitting the prompt
again---and hence directly translated into a lack of voice, agency, and dignity.
Yet, as stated in the second sentence, the alleged policy violation was
considered so severe that it got close to warranting deplatforming. The claims
of excessive severity and attendant punitive threat were very much unexpected
and rather unreasonable given Bacon's prominent stature in fine art. They
further denied voice, agency, and dignity. In addition to each sentence
representing an injustice by itself, the epistemic dissonance between the two
sentences made the warning's overall injustice more acute---and thereby also
more noticeable.

The warning's first sentence did link to \DALLE's content policy, which is
reproduced in \REF{app:dalle-content-policy}. It started with this general
command:

\begin{quote}
Do not attempt to create, upload, or share images that are not G-rated or that
could cause harm.
\end{quote}

\noindent{}That command was followed by a hierarchy of strictures that
elaborated on this apparent prime directive. They included generally desirable
strictures, notably requiring the disclosure of \V{AI} and respect for the
rights of others. But the combination of two other strictures with a clause from
the addendum to OpenAI's terms-of-use, as reproduced in \REF{app:dalle:terms},
seemed less concerned with preventing harm than taking control in depth:

\begin{EnumerateLining}
\item OpenAI presented the categories of violative content as a prescriptive
    jumble of nouns and adjectives, without justification or explanation:
    ``Hate,'' ``harassment,'' ``violence,'' ``self-harm,'' ``sexual,''
    ``shocking,'' ``illegal activity,'' ``deception,'' ``political,'' ``public
    and personal health,'' and ``spam.''
\item OpenAI claimed ownership to all generated images, instead granting
    exclusive usage rights only---``all provided that you comply with these
    terms and our Content Policy.'' The addendum to the terms-of-use continued:
    ``If you violate our terms or Content Policy, you will lose rights to use
    Generations,'' meaning images.
\item OpenAI invited users to ``report any suspected violations,'' i.e., to
    become snitches, and promised to ``take action accordingly, up to and
    including terminating the violating account.''
\end{EnumerateLining}

\DALLE's \V{AI}-based prompt moderation was preemptive and based on prescriptive
and overreaching content categories. By omitting motivation or justification,
the categories did not allow for context or marginal speech---preventing users
from making informed decisions about whether prompts adhered to the policy.
Pretty much their only safe option was to stay well clear of prompts that might
touch upon the eleven categories. Furthermore, while it is straight-forward
enough to come up with examples for potentially harmful content in each
category, declaring all violent, sexual, shocking, political, or health content
harmful seems rather preposterous.

By contrast, retaining ownership and encouraging users to snitch are only
relevant for post-hoc enforcement of \DALLE's policy, albeit at a much slower
pace and with significantly more cost and effort. Since \V{AI}-generated imagery
is not a human creation, it also isn't copyrightable in the United States. That
immediately puts instruments, such as \V{DMCA} takedown notices and the
Copyright Claims Board~\cite{CCB2022}, out of reach. The former helps with
removing copyrighted content hosted by American firms and the latter makes for
much faster decisions than the courts. As added bonus, neither requires a
lawyer. Instead, OpenAI would have to rely on contract law and seek a court
order instead, suggesting that these two strictures were for highly visible,
reputational threats only. They certainly would have been no match for the sheer
volume of images produced by \DALLE: 2 million images per day by late September
2022~\cite{OpenAI2022a} and over 4 million per day by early November
2022~\cite{OpenAI2022h}!

Seen through the lens of procedural justice, OpenAI's content policy and
terms-of-use sought to impose exceedingly broad prohibitions on its users
without justification, depriving them of their voice. Their aggressive,
automated enforcement stripped users of agency to make their own decisions about
what content is appropriate in what context and instead treated them with
punitive contempt. If that wasn't enough, OpenAI retained ownership rights,
licensing image use only, and encouraged users to serve as informants, impinging
on their dignity.

\subsection{OpenAI's Neutrality and Trustworthiness}

Before \DALLE~2's beta opened up the system to users like me, it was available
to a much smaller number of users, more like a research experiment. The system
card~\cite{GreenProcopeea2022,ProcopeCheemaea2022} for the original release in
April 2022 makes clear that \DALLE~2 was, in part, trained with ``publicly
available sources''~\cite{MishkinAhmad2022}, which in all likelihood included
Internet-sourced data similar to the \V{LAION-400M}
dataset~\cite{SchuhmannVencuea2021}. While OpenAI has declined to elaborate on
the exact sources for \DALLE's training data, we know that such Internet-sourced
datasets are anything but safe~\cite{BirhanePrabhuea2021}. That makes \DALLE\
unsafe by design.

As shown in \REF{app:dalle-content-policy}, \DALLE's content policy came right
out against anything that is ``not G-rated'' or ``that could cause harm.'' In
their \V{FAQ} entry for \DALLE's warnings~\cite{Natalie2022}, OpenAI claimed
that ``safe usage of the platform is our highest priority.'' While some of the
violative categories include classes of content known to be causing human
harm---e.g., exposure to violent games~\cite{AndersonShibuyaea2010,
PrescottSargentea2018} or sexual content~\cite{MoriParkea2023,
RodenhizerEdwards2019} for children and adolescents as well as exposure to
political, health-related, and other misinformation for society in
general~\cite{DennissLindberg2025, VasistChatterjeeea2024}, but also
see~\cite{AdamsOsmanea2023}---\DALLE's content policy disallowed \emph{all}
violent, sexual, political, and health content. Especially the prohibitions
against the latter two categories are not just unusually broad, they also run
directly counter the public interest in a democracy. At the same time, health,
like politics over the years before, has become an exceedingly partisan topic
during the pandemic. That suggests a very specific kind of harm OpenAI is
seeking protection from---harm to its own reputation.

When OpenAI gave up control over generations in early November 2022, the email
announcement justified that change with ``improvements in our safety systems.''
That may be the case for generations created after those improvements were made.
But when I asked customer support about generations made before the
announcement, they confirmed that the new terms-of-use ``apply to all
generations, regardless of the date on which they were made.'' In fact, OpenAI
deleted the webpage with the terms-of-use addendum for \DALLE\ in November 2022.
But if \DALLE's \V{AI}-based content moderation required improvements, then
chances are that at least some older images were unsafe. Otherwise, there would
have been no need for those safety improvements. Yet OpenAI pretended it can
have it both ways.

One possible explanation for this contradictory stance is market pressure
stemming from the August~2022 release of Stable Diffusion, a competing
text-to-image model created by Stability~\V{AI}~\cite{StabilityAI2022}. Unlike
OpenAI with \DALLE, Stability~\V{AI} released the source code and model weights
for Stable Diffusion. That enabled anyone with basic fluency in Python and
access to recent graphics cards by Nvidia to open their own competitor to
\DALLE, without paying a license fee to Stability~\V{AI} and without OpenAI's
restrictive content policy. While that is pure speculation on my part,
Microsoft's \$10 billion investment in OpenAI in January 2023 on top of an
earlier \$1 billion investment illustrates the stakes at play~\cite{Bass2023},
which are a powerful motivation to cut corners.

In summary, OpenAI was exceedingly strategical about policies and information
released to the public. Alas, the firm's apparent need for controlling
everything \DALLE\ was overbearing, bordering on the arrogant or patrician, even
though some of its positions are blatantly hypocritical. These impressions seem
in line with Karen Hao's observations in a 2020 portrait of the
firm~\cite{Hao2020}, which focused on the tensions between OpenAI's founding as
a research lab and its current increasingly commercial activities. However, more
recent reporting about the firm's use of labor in Latin America and Africa for
labelling content is alarming~\cite{HaoHernandez2022,Perrigo2023a}. If
confirmed, OpenAI engaged in outsourcing practices that directly harmed people
and probably also violated \V{US} law. That raises grave concerns about the
firm's ethics and its ability to follow through on its ambitious
charter~\cite{OpenAI2018}.
 \section{March 2024: Conflict of Pinterest}
\label{sec:pinterest}

I've been maintaining a fetish-themed picture collection (``board'') on
Pinterest since the winter of 2018/19. While I uploaded a couple of new images,
the board mostly collects images already present on Pinterest, none of them
sexually explicit. As part of what appeared to be a sweep through existing
content with a new machine learning model, Pinterest removed five of the images
(``pins'') over four days in March 2024. \REF{app:pinterest:notification}
reproduces the text of the corresponding email notifications.

Pinterest's emails clearly stated the reason for the take-downs, repeating
relevant language from its community guidelines. Compared to Twitter, the
guidelines are expressed far more succinctly, for instance, lacking detailed
examples. Still, in case of my pins, the reason for their take-down was clear
enough, as they all were ``fetish imagery.''

Pinterest's notifications also appeared to link to the offending content, an
important affordance when retro-actively moderating images. Alas, that link was
a one-time link and only available for seven days after Pinterest sent out the
notification. On top of that, linked images were not the originals but severely
blurred versions thereof. That rendered some of them unrecognizable and all of
them unrecoverable, thus preventing me from archiving the material on my own
computer.

Pinterest providing at-most-once access to a blurred version of the original
image stands in stark contrast to the receiver of the take-down notice being the
content curator, who (presumably) not only reviewed the image but also approved
of it. Meanwhile, Pinterest's overly short deadline for inspecting the blurred
version stands in stark contrast to the firm publicly hosting the original image
for several years. On its own, Pinterest blurring images already seems unduly
precious. When combined with at-most-once and time-gated links, the firm appears
to manufacture a sense of transgressive severity (``we can't possibly expose you
to such content!'') and restorative urgency (``we must purge that content from
our systems \V{ASAP}!''). Such manipulation may be motivated by a desire to
impart a lesson or instill compliance. But it also violates users' voice,
agency, as well as dignity and hence represents the first injustice.

Just as for the previous two case studies, the threat of banishment was
prominently stated in Pinterest's notification email. Like Twitter and OpenAI,
Pinterest provided no information on the exact conditions for account suspension
or termination---neither in the notification email nor in linked policies. Since
my account remains active to this day, five violative images, none uploaded by
me, appear to be acceptable. Yet, algorithmic enforcement of the stated policy
remains partial, as my board continues to feature similar images---the one
difference being that remaining images display fetish implements by themselves
and not \emph{in situ}, on human bodies. This inconsistent and unpredictable
enforcement, under persistent threat of banishment, further denies users their
voice as well as agency and represents the second injustice.

\subsection{Pinterest's Neutrality and Trustworthiness}

Perusal of discussions on Reddit's r/Pinterest forum in early March
2024~\cite{ArmKooky2024, itsmevic1112024, Jacedayton2024} demonstrates that I am
far from the only user, whose pins were taken down \emph{en masse} that month.
While my pins were, in fact, violating Pinterest's stated policy, many
comments on r/Pinterest complained about false positives, e.g., pictures of
classical paintings, tattoos, or fashion shows being taken down. However, at
least paintings and tattoos should qualify for the exemption of
``nudity in paintings and sculptures and in science and historical contexts'' in
Pinterest's community guidelines (see \REF{app:pinterest:policy}). A few
comments also claimed that their accounts were terminated, including when only
re-posting already existing pins; unfortunately, they did not include the number
of violative images and thus failed to help narrow down Pinterest's actual
threshold for account termination.

At the time, Pinterest provided no public guidance on its more aggressive
enforcement of prohibitions against adult content. It acknowledged the
\V{Q1}~2024 deployment of a machine learning model for weeding out adult content
only well over a year later in its transparency report
for the first half of 2024: ``In \V{Q1}~2024 we broadened our use of automated
tools to deactivate content for [showing adult content]''~\cite{Pinterest2025}.
The accompanying data, however, suggests that fully automated enforcement---with
the \V{AI} acting as judge, jury, and executioner---was short-lived for the most
part, comprising 0\% of all deactivations for adult content in \V{Q3}~2023, <1\%
in \V{Q4}~2023, 37\% in \V{Q1}~2024, and 3\% in \V{Q2}~2024. It also suggests
that accuracy suffered, with the fraction of successful appeals for adult
content increasing from 8\% in \V{Q3}~2023 and 18\% in \V{Q4}~2023 to 43\% in
\V{Q1}~2024 and 33\% in \V{Q2}~2024.

Given Meta's well-documented troubles with over-moderating naked human bodies on
Facebook and Instagram~\cite{Gillespie2018, Ortutay2020,
StjernfeltLauritzen2019}, it shouldn't take hindsight to realize that Pinterest
switching from mostly manual to \V{AI}-based moderation of adult content would
run into challenges. In other words, the firm would have been well advised to
more carefully tune its stochastic detection model and to more proactively
communicate its updated policy enforcement, including in notification emails.
Instead, Pinterest committed engineering resources to implementing web-based,
at-most-once access to blurred images. It's unclear whether Pinterest's
implementation is sufficiently robust---i.e., implements at-most-once semantics
end-to-end instead of only for \V{HTTP} requests~\cite{SaltzerReedea1984}---and
scalable---i.e., deduplicates storage for the 7.4 pins sharing the same image
for adult content during \V{Q1}~2024. At the same time, a much simpler delivery
mechanism has been readily available to the firm's engineers: Including
violative images in notification emails. If images were left unblurred, users
could (trivially) recover their content as well.

Pinterest's apparent prioritization of a mechanism to boost impressions of
severity and urgency over more accurate detection of violative content and
transparent communication of enforcement actions raises significant doubts about
the firm's neutrality and trustworthiness. These doubts are reinforced by
Pinterest eliminating the above quoted exception for nudity in art, science, and
history in June 2025, at least for the \V{US}~\cite{Schneider2025}. It amounts
to a tacit admission that the firm is struggling with the moderation of adult
content in the large. While banning all nudity may seemingly address these
challenges, it directly depresses platform quality and ends up alienating users,
by unreasonably limiting their speech and by banishing them to the stochastic
penal colony.

As illustrated by my experience with Pinterest's recommendation algorithm when I
first started using the platform during the winter of 2018/2019, effective
moderation of sexual(ized) content needs to accommodate the many ways humans
bestow (sexual) meaning to images. I was exploring sports-themed images of men,
when Pinterest's algorithm started mixing images of similarly dressed, barely
teenage boys into the results. Boys like men were fully dressed, so the images
did not violate even a total ban on nudity. Additionally, individual images were
not sexualized, at least overtly. It was only their aggregation under a
fetishistic gaze that produced clear sexual connotations---and therefore made
the inclusion of boys in search results highly inappropriate. It took some
persistence to reach a customer service representative for Pinterest, who
promised to alert the responsible engineering team. Thankfully, I could not
reproduce this behavior a few months later. Still, the episode serves as
reminder that not all meaning is overt and that implicit, indirect, and
contextual meaning is much harder to detect and moderate, especially
algorithmically.
 \section{Conclusion}
\label{sec:conclusion}

This paper provided three partial maps of the stochastic penal colony, i.e., the
punishment processes resulting from algorithmic interventions, notably content
moderation by Twitter in October 2021, OpenAI's \DALLE\ in July 2022, and
Pinterest in March 2024. Common to all three are underspecified threats of
banishment for posting/generating violative content, which is reviving the
historical practice of transportation for the digital age. Otherwise, the three
three differ substantially, with Twitter staging a demeaning punishment
performance rooted in an effective othering, OpenAI going as far as encouraging
denunciation to control its users, and Pinterest manufacturing violative
severity as well as urgency instead of implementing more accurate detection and
transparent notification processes.

What is missing is an acknowledgement by the case study organizations that
content moderation in the large simply cannot be accurate or
fair~\cite{Masnick2019}. It seems positively futile when considering that even
the Hebrew Bible, which serves as foundational text to three major world
religions, includes plenty of (aberrant) sex and violence. For example,
Genesis~19 features all male inhabitants of Sodom (save Lot) clamoring to gang
rape \YHWH's messengers, \YHWH\ destroying the cities of Sodom and Gomorrah,
also of Admah and Zeboim, their populations, and surrounding fields with a rain
of fire and brimstone, Lot's wife being turned into a salt pillar for copping a
glance at \YHWH\ during Their orgy of destruction, and Lot's surviving two
virgin daughters successfully conspiring to get daddy drunk so that he knocks
them up.

In fact, by observing prompts and images shared on Reddit's \DALLE~2 forum for
several months following July 2022, we could eliminate other candidate
categories and determine that Biblical violence, more precisely the violence of
the crucifixion, triggered the prompt moderation in \REF{sec:dalle}.
Interestingly, OpenAI added an exemption for crucifixion to \DALLE\ some time
before February 2023, at which time the prompt was not rejected anymore.

Meanwhile current events in the United States---with the Trump administration
weaponizing the Department of Justice and \V{FBI} for political
retribution~\cite{BazelonPoser2025, BazelonPoser2026}, arresting and killing
citizens during brutal immigration crackdowns~\cite{Barajas2026, Maney2025,
Sandoval2026}, subsuming public health to dangerous
ideology~\cite{Interlandi2026}, systematically attacking
academia~\cite{DuBozeman2025, Owen-Smith2025}, banning people from visiting the
country for criticizing them~\cite{Mackey2025,Myers2025}, and so on---point
towards the terrifying possibility that variations of punishment as performance
and the literal penal colony may become instruments of government oppression in
the largest and most powerful Western democracy. One can only wonder about the
role of social media's punishment practices in preparing us for this turn away
from liberal democracy. Welcome to the penal colony~\emoji{desert-island},
stochastic or otherwise!
 
\section*{Generative AI Disclosure Statement}

This paper was conceptualized and written entirely \emph{without} the help of
generative \V{AI}.

\begin{acks}
We thank the anonymous FAccT reviewers for their detailed and helpful feedback
on earlier versions of this paper. We also thank Cordula Hahn, David Halperin,
Akiko Kyei-Aboagye, Thomas Schr\"{o}ter, Karin Wolman, and Petra Zaus for their
help with conceptualizing and presenting this material.

This work was supported in part by \V{MEYS}, \V{ERC} \V{CZ} program, grant
\V{LL-2325}.
\end{acks}

\bibliographystyle{ACM-Reference-Format}
\bibliography{bibliography}


\begin{thebibliography}{136}


\ifx \showCODEN    \undefined \def \showCODEN     #1{\unskip}     \fi
\ifx \showDOI      \undefined \def \showDOI       #1{#1}\fi
\ifx \showISBNx    \undefined \def \showISBNx     #1{\unskip}     \fi
\ifx \showISBNxiii \undefined \def \showISBNxiii  #1{\unskip}     \fi
\ifx \showISSN     \undefined \def \showISSN      #1{\unskip}     \fi
\ifx \showLCCN     \undefined \def \showLCCN      #1{\unskip}     \fi
\ifx \shownote     \undefined \def \shownote      #1{#1}          \fi
\ifx \showarticletitle \undefined \def \showarticletitle #1{#1}   \fi
\ifx \showURL      \undefined \def \showURL       {\relax}        \fi
\providecommand\bibfield[2]{#2}
\providecommand\bibinfo[2]{#2}
\providecommand\natexlab[1]{#1}
\providecommand\showeprint[2][]{arXiv:#2}

\bibitem[{Access Now} et~al\mbox{.}(2021)]%
        {AccessNowACLUFoundationOfNorthernCaliforniaea2021}
\bibfield{author}{\bibinfo{person}{{Access Now}}, \bibinfo{person}{{ACLU
  Foundation of Northern California}}, \bibinfo{person}{{ACLU Foundation of
  Southern California}}, \bibinfo{person}{{Article 19}},
  \bibinfo{person}{{Brennan Center for Justice}}, \bibinfo{person}{{Center for
  Democracy \& Technology}}, \bibinfo{person}{{Electronic Frontier
  Foundation}}, \bibinfo{person}{{Global Partners Digital}},
  \bibinfo{person}{{InternetLab}}, \bibinfo{person}{{National Coalition Against
  Censorship}}, \bibinfo{person}{{New America's Open Technology Institute}},
  \bibinfo{person}{{Ranking Digital Rights}}, \bibinfo{person}{{Red en Defensa
  de los Derechos Digitales}}, {and} \bibinfo{person}{{Witness}}.}
  \bibinfo{year}{2021}\natexlab{}.
\newblock \bibinfo{title}{Santa {{Clara Principles}} on {{Transparency}} and
  {{Accountability}} in {{Content Moderation}}}.
\newblock
\newblock
\urldef\tempurl%
\url{https://santaclaraprinciples.org}
\showURL{%
\tempurl}


\bibitem[Adams et~al\mbox{.}(2023)]%
        {AdamsOsmanea2023}
\bibfield{author}{\bibinfo{person}{Zo{\"e} Adams}, \bibinfo{person}{Magda
  Osman}, \bibinfo{person}{Christos Bechlivanidis}, {and}
  \bibinfo{person}{Bj{\"o}rn Meder}.} \bibinfo{year}{2023}\natexlab{}.
\newblock \showarticletitle{({{Why}}) {{Is Misinformation}} a {{Problem}}?}
\newblock \bibinfo{journal}{\emph{Perspectives on Psychological Science}}
  \bibinfo{volume}{18}, \bibinfo{number}{6} (\bibinfo{date}{Feb.}
  \bibinfo{year}{2023}), \bibinfo{pages}{1436--1463}.
\newblock
\urldef\tempurl%
\url{https://doi.org/10.1177/17456916221141344}
\showURL{%
\tempurl}


\bibitem[Aldrich(2010)]%
        {Aldrich2010}
\bibfield{author}{\bibinfo{person}{Robert Aldrich}.}
  \bibinfo{year}{2010}\natexlab{}.
\newblock \showarticletitle{The {{French Overseas Empire}} and Its
  {{Contemporary Legacy}}}.
\newblock \bibinfo{journal}{\emph{European History Quarterly}}
  \bibinfo{volume}{40}, \bibinfo{number}{1} (\bibinfo{date}{Jan.}
  \bibinfo{year}{2010}), \bibinfo{pages}{97--108}.
\newblock
\showISSN{0265-6914}
\urldef\tempurl%
\url{https://doi.org/10.1177/0265691409351339}
\showDOI{\tempurl}


\bibitem[Anderson(2018)]%
        {Anderson2018}
\bibfield{editor}{\bibinfo{person}{Clare Anderson}} (Ed.).
  \bibinfo{year}{2018}\natexlab{}.
\newblock \bibinfo{booktitle}{\emph{A {{Global History}} of {{Convicts}} and
  {{Penal Colonies}}}}.
\newblock \bibinfo{publisher}{Bloomsbury Academic}, \bibinfo{address}{New York,
  NY, USA}.
\newblock
\urldef\tempurl%
\url{https://doi.org/10.5040/9781350000704}
\showDOI{\tempurl}


\bibitem[Anderson et~al\mbox{.}(2010)]%
        {AndersonShibuyaea2010}
\bibfield{author}{\bibinfo{person}{Craig~A. Anderson}, \bibinfo{person}{Akiko
  Shibuya}, \bibinfo{person}{Nobuko Ihori}, \bibinfo{person}{Edward~L. Swing},
  \bibinfo{person}{Brad~J. Bushman}, \bibinfo{person}{Akira Sakamoto},
  \bibinfo{person}{Hannah~R. Rothstein}, {and} \bibinfo{person}{Muniba
  Saleem}.} \bibinfo{year}{2010}\natexlab{}.
\newblock \showarticletitle{Violent Video Game Effects on Aggression, Empathy,
  and Prosocial Behavior in {{Eastern}} and {{Western}} Countries: {{A}}
  Meta-Analytic Review.}
\newblock \bibinfo{journal}{\emph{Psychological Bulletin}}
  \bibinfo{volume}{136}, \bibinfo{number}{2} (\bibinfo{year}{2010}),
  \bibinfo{pages}{151--173}.
\newblock
\showISSN{1939-1455, 0033-2909}
\urldef\tempurl%
\url{https://doi.org/10.1037/a0018251}
\showDOI{\tempurl}


\bibitem[Angwin et~al\mbox{.}(2016)]%
        {AngwinLarsonea2016}
\bibfield{author}{\bibinfo{person}{Julia Angwin}, \bibinfo{person}{Jeff
  Larson}, \bibinfo{person}{Surya Mattu}, {and} \bibinfo{person}{Lauren
  Kirchner}.} \bibinfo{year}{2016}\natexlab{}.
\newblock \showarticletitle{Machine {{Bias}}}.
\newblock \bibinfo{journal}{\emph{ProPublica}} (\bibinfo{date}{May}
  \bibinfo{year}{2016}).
\newblock
\urldef\tempurl%
\url{https://www.propublica.org/article/machine-bias-risk-assessments-in-criminal-sentencing}
\showURL{%
\tempurl}


\bibitem[{Anonymous}(2016)]%
        {Anonymous2016}
\bibfield{author}{\bibinfo{person}{{Anonymous}}.}
  \bibinfo{year}{2016}\natexlab{}.
\newblock \showarticletitle{Incident 37: {{Female Applicants Down-Ranked}} by
  {{Amazon Recruiting Tool}}}.
\newblock \bibinfo{journal}{\emph{Artificial Intelligence Incident Database}}
  (\bibinfo{date}{Aug.} \bibinfo{year}{2016}).
\newblock
\urldef\tempurl%
\url{https://incidentdatabase.ai/cite/37/}
\showURL{%
\tempurl}


\bibitem[{Anonymous}(2018)]%
        {Anonymous2018}
\bibfield{author}{\bibinfo{person}{{Anonymous}}.}
  \bibinfo{year}{2018}\natexlab{}.
\newblock \showarticletitle{Incident 405: {{Schufa Credit Scoring}} in
  {{Germany Reported}} for {{Unreliable}} and {{Imbalanced Scores}}}.
\newblock \bibinfo{journal}{\emph{Artificial Intelligence Incident Database}}
  (\bibinfo{date}{Nov.} \bibinfo{year}{2018}).
\newblock
\urldef\tempurl%
\url{https://incidentdatabase.ai/cite/405}
\showURL{%
\tempurl}


\bibitem[Are(2023)]%
        {Are2023}
\bibfield{author}{\bibinfo{person}{Carolina Are}.}
  \bibinfo{year}{2023}\natexlab{}.
\newblock \showarticletitle{An Autoethnography of Automated Powerlessness:
  Lacking Platform Affordances in {{Instagram}} and {{TikTok}} Account
  Deletions}.
\newblock \bibinfo{journal}{\emph{Media, Culture \& Society}}
  \bibinfo{volume}{45}, \bibinfo{number}{4} (\bibinfo{date}{May}
  \bibinfo{year}{2023}), \bibinfo{pages}{822--840}.
\newblock
\showISSN{0163-4437, 1460-3675}
\urldef\tempurl%
\url{https://doi.org/10.1177/01634437221140531}
\showDOI{\tempurl}


\bibitem[{ArmKooky}(2024)]%
        {ArmKooky2024}
\bibfield{author}{\bibinfo{person}{{ArmKooky}}.}
  \bibinfo{year}{2024}\natexlab{}.
\newblock \bibinfo{title}{Pinterest's {{Unjust Content Moderation}}}.
\newblock
\newblock
\urldef\tempurl%
\url{https://web.archive.org/web/20250327185105/https://www.reddit.com/r/Pinterest/comments/1b9tzsd/pinterests_unjust_content_moderation/}
\showURL{%
\tempurl}


\bibitem[Arnsdorf(2020)]%
        {Arnsdorf2020}
\bibfield{author}{\bibinfo{person}{Isaac Arnsdorf}.}
  \bibinfo{year}{2020}\natexlab{}.
\newblock \showarticletitle{Inside {{Trump}} and {{Barr}}'s {{Last-Minute
  Killing Spree}}}.
\newblock \bibinfo{journal}{\emph{ProPublica}} (\bibinfo{date}{Dec.}
  \bibinfo{year}{2020}).
\newblock
\urldef\tempurl%
\url{https://www.propublica.org/article/inside-trump-and-barrs-last-minute-killing-spree/amp}
\showURL{%
\tempurl}


\bibitem[Atherton(2022)]%
        {Atherton2022a}
\bibfield{author}{\bibinfo{person}{Daniel Atherton}.}
  \bibinfo{year}{2022}\natexlab{}.
\newblock \showarticletitle{Incident 303: {{Google}}'s {{Automated Child Abuse
  Detection Wrongfully Flagged}} a {{Parent}}'s {{Naked Photo}} of {{His
  Child}}}.
\newblock \bibinfo{journal}{\emph{Artificial Intelligence Incident Database}}
  (\bibinfo{date}{Aug.} \bibinfo{year}{2022}).
\newblock
\urldef\tempurl%
\url{https://incidentdatabase.ai/cite/303}
\showURL{%
\tempurl}


\bibitem[Barajas(2026)]%
        {Barajas2026}
\bibfield{author}{\bibinfo{person}{Joshua Barajas}.}
  \bibinfo{year}{2026}\natexlab{}.
\newblock \bibinfo{title}{Shooting Deaths Climb in {{Trump}}'s Mass Deportation
  Effort}.
\newblock
\newblock
\urldef\tempurl%
\url{https://www.pbs.org/newshour/nation/a-look-at-shootings-by-federal-immigration-officers}
\showURL{%
\tempurl}


\bibitem[Bass(2023)]%
        {Bass2023}
\bibfield{author}{\bibinfo{person}{Dina Bass}.}
  \bibinfo{year}{2023}\natexlab{}.
\newblock \showarticletitle{Microsoft {{Invests}} \$10 {{Billion}} in {{ChatGPT
  Maker OpenAI}}}.
\newblock \bibinfo{journal}{\emph{Bloomberg}} (\bibinfo{date}{Jan.}
  \bibinfo{year}{2023}).
\newblock
\urldef\tempurl%
\url{https://www.bloomberg.com/news/articles/2023-01-23/microsoft-makes-multibillion-dollar-investment-in-openai}
\showURL{%
\tempurl}


\bibitem[Bazelon and Poser(2025)]%
        {BazelonPoser2025}
\bibfield{author}{\bibinfo{person}{Emily Bazelon} {and} \bibinfo{person}{Rachel
  Poser}.} \bibinfo{year}{2025}\natexlab{}.
\newblock \showarticletitle{`{{It}}'s a {{Culture Now}} of {{Fear}}': {{A
  Year}} of {{Chaos Inside}} the {{Justice Department}}}.
\newblock \bibinfo{journal}{\emph{The New York Times}} (\bibinfo{date}{Nov.}
  \bibinfo{year}{2025}).
\newblock
\showISSN{0362-4331}
\urldef\tempurl%
\url{https://www.nytimes.com/interactive/2025/11/16/magazine/trump-justice-department-staff-attorneys.html}
\showURL{%
\tempurl}


\bibitem[Bazelon and Poser(2026)]%
        {BazelonPoser2026}
\bibfield{author}{\bibinfo{person}{Emily Bazelon} {and} \bibinfo{person}{Rachel
  Poser}.} \bibinfo{year}{2026}\natexlab{}.
\newblock \showarticletitle{Inside {{Kash Patel}}'s {{F}}.{{B}}.{{I}}.:
  {{Meltdowns}}, {{Chaos}}, {{Vendettas}}}.
\newblock \bibinfo{journal}{\emph{The New York Times}} (\bibinfo{date}{Jan.}
  \bibinfo{year}{2026}).
\newblock
\showISSN{0362-4331}
\urldef\tempurl%
\url{https://www.nytimes.com/interactive/2026/01/22/magazine/trump-kash-patel-fbi-agents.html}
\showURL{%
\tempurl}


\bibitem[Bender et~al\mbox{.}(2021)]%
        {BenderGebruea2021}
\bibfield{author}{\bibinfo{person}{Emily~M. Bender}, \bibinfo{person}{Timnit
  Gebru}, \bibinfo{person}{Angelina {McMillan-Major}}, {and}
  \bibinfo{person}{Shmargaret Shmitchell}.} \bibinfo{year}{2021}\natexlab{}.
\newblock \showarticletitle{On the {{Dangers}} of {{Stochastic Parrots}}: {{Can
  Language Models Be Too Big}}? \emo{parrot}}. In
  \bibinfo{booktitle}{\emph{Proceedings of the 2021 {{ACM Conference}} on
  {{Fairness}}, {{Accountability}}, and {{Transparency}}}}
  \emph{(\bibinfo{series}{{{FAccT}} '21})}. \bibinfo{publisher}{Association for
  Computing Machinery}, \bibinfo{address}{Virtual Event, Canada},
  \bibinfo{pages}{610--623}.
\newblock
\showISBNx{978-1-4503-8309-7}
\urldef\tempurl%
\url{https://doi.org/10.1145/3442188.3445922}
\showDOI{\tempurl}


\bibitem[Birhane et~al\mbox{.}(2021)]%
        {BirhanePrabhuea2021}
\bibfield{author}{\bibinfo{person}{Abeba Birhane}, \bibinfo{person}{Vinay~Uday
  Prabhu}, {and} \bibinfo{person}{Emmanuel Kahembwe}.}
  \bibinfo{year}{2021}\natexlab{}.
\newblock \bibinfo{title}{Multimodal Datasets: Misogyny, Pornography, and
  Malignant Stereotypes}.
\newblock
\newblock
\showeprint[arxiv]{2110.01963}~[cs]
\urldef\tempurl%
\url{http://arxiv.org/abs/2110.01963}
\showURL{%
\tempurl}


\bibitem[Brown(2019)]%
        {Brown2019a}
\bibfield{author}{\bibinfo{person}{H.~Claire Brown}.}
  \bibinfo{year}{2019}\natexlab{}.
\newblock \showarticletitle{Amazon's {{On-Site Emergency Care Endangers Its Own
  Employees}}}.
\newblock \bibinfo{journal}{\emph{The Intercept}} (\bibinfo{date}{Dec.}
  \bibinfo{year}{2019}).
\newblock
\urldef\tempurl%
\url{https://theintercept.com/2019/12/02/amazon-warehouse-workers-safety-cyber-monday/}
\showURL{%
\tempurl}


\bibitem[Bucher(2018)]%
        {Bucher2018}
\bibfield{author}{\bibinfo{person}{Taina Bucher}.}
  \bibinfo{year}{2018}\natexlab{}.
\newblock \bibinfo{booktitle}{\emph{If...{{Then}}: {{Algorithmic Power}} and
  {{Politics}}}}.
\newblock \bibinfo{publisher}{Oxford University Press},
  \bibinfo{address}{Oxford, United Kingdom}.
\newblock
\urldef\tempurl%
\url{10.1093/oso/9780190493028.001.0001}
\showURL{%
\tempurl}


\bibitem[Casovan(2022)]%
        {Casovan2022}
\bibfield{author}{\bibinfo{person}{Ashley Casovan}.}
  \bibinfo{year}{2022}\natexlab{}.
\newblock \showarticletitle{Incident 430: {{Lawyers Denied Entry}} to
  {{Performance Venue}} by {{Facial Recognition}}}.
\newblock \bibinfo{journal}{\emph{Artificial Intelligence Incident Database}}
  (\bibinfo{date}{Dec.} \bibinfo{year}{2022}).
\newblock
\urldef\tempurl%
\url{https://incidentdatabase.ai/cite/430}
\showURL{%
\tempurl}


\bibitem[{CCB}(2022)]%
        {CCB2022}
\bibfield{author}{\bibinfo{person}{{CCB}}.} \bibinfo{year}{2022}\natexlab{}.
\newblock \bibinfo{title}{Copyright {{Claims Board Handbook}}}.
\newblock
\newblock
\urldef\tempurl%
\url{https://ccb.gov/handbook/}
\showURL{%
\tempurl}


\bibitem[Chughtai(2022)]%
        {Chughtai2022}
\bibfield{author}{\bibinfo{person}{Alia Chughtai}.}
  \bibinfo{year}{2022}\natexlab{}.
\newblock \showarticletitle{Mapping the Scale of Damage by the Catastrophic
  {{Pakistan}} Floods \textbar{} {{Infographic News}} \textbar{} {{Al
  Jazeera}}}.
\newblock \bibinfo{journal}{\emph{Al Jazeera}} (\bibinfo{date}{Sept.}
  \bibinfo{year}{2022}).
\newblock
\urldef\tempurl%
\url{https://www.aljazeera.com/news/longform/2022/9/16/mapping-the-scale-of-destruction-of-the-pakistan-floods}
\showURL{%
\tempurl}


\bibitem[Clark(2023)]%
        {Clark2023}
\bibfield{author}{\bibinfo{person}{Mitchell Clark}.}
  \bibinfo{year}{2023}\natexlab{}.
\newblock \showarticletitle{Amazon's {{OSHA}} Fine for Warehouse Safety
  Violations Could Be about \${{60K}}}.
\newblock \bibinfo{journal}{\emph{The Verge}} (\bibinfo{date}{Jan.}
  \bibinfo{year}{2023}).
\newblock
\urldef\tempurl%
\url{https://www.theverge.com/2023/1/18/23561506/amazon-osha-citations-ergonomics-struck-by-pace}
\showURL{%
\tempurl}


\bibitem[Covert(2022)]%
        {Covert2022}
\bibfield{author}{\bibinfo{person}{Bryce Covert}.}
  \bibinfo{year}{2022}\natexlab{}.
\newblock \showarticletitle{The {{Little-Known Policy Wreaking Havoc}} on
  {{Workers}}' {{Lives}}}.
\newblock \bibinfo{journal}{\emph{Intelligencer}} (\bibinfo{date}{Aug.}
  \bibinfo{year}{2022}).
\newblock
\urldef\tempurl%
\url{https://nymag.com/intelligencer/2022/08/the-little-known-policy-wreaking-havoc-on-workers-lives.html}
\showURL{%
\tempurl}


\bibitem[Dave(2022)]%
        {Dave2022}
\bibfield{author}{\bibinfo{person}{Amit Dave}.}
  \bibinfo{year}{2022}\natexlab{}.
\newblock \showarticletitle{Birds Fall from the Sky as Heatwave Scorches
  {{India}}}.
\newblock \bibinfo{journal}{\emph{Reuters}} (\bibinfo{date}{May}
  \bibinfo{year}{2022}).
\newblock
\urldef\tempurl%
\url{https://www.reuters.com/world/india/birds-fall-sky-heatwave-scorches-india-2022-05-11/}
\showURL{%
\tempurl}


\bibitem[Dehlendorf and Gerety(2021)]%
        {DehlendorfGerety2021}
\bibfield{author}{\bibinfo{person}{Andrea Dehlendorf} {and}
  \bibinfo{person}{Ryan Gerety}.} \bibinfo{year}{2021}\natexlab{}.
\newblock \showarticletitle{The {{Punitive Potential}} of {{AI}}}.
\newblock \bibinfo{journal}{\emph{Boston Review}} (\bibinfo{date}{May}
  \bibinfo{year}{2021}).
\newblock
\urldef\tempurl%
\url{https://www.bostonreview.net/forum_response/the-punitive-potential-of-ai/}
\showURL{%
\tempurl}


\bibitem[Denniss and Lindberg(2025)]%
        {DennissLindberg2025}
\bibfield{author}{\bibinfo{person}{Emily Denniss} {and}
  \bibinfo{person}{Rebecca Lindberg}.} \bibinfo{year}{2025}\natexlab{}.
\newblock \showarticletitle{Social Media and the Spread of Misinformation:
  Infectious and a Threat to Public Health}.
\newblock \bibinfo{journal}{\emph{Health Promotion International}}
  \bibinfo{volume}{40}, \bibinfo{number}{2} (\bibinfo{date}{March}
  \bibinfo{year}{2025}), \bibinfo{pages}{daaf023}.
\newblock
\showISSN{0957-4824, 1460-2245}
\urldef\tempurl%
\url{https://doi.org/10.1093/heapro/daaf023}
\showDOI{\tempurl}


\bibitem[Drake(2005)]%
        {Drake2005}
\bibfield{author}{\bibinfo{person}{David Drake}.}
  \bibinfo{year}{2005}\natexlab{}.
\newblock \showarticletitle{The {{Dreyfus Affair}} and the {{Birth}} of the
  `{{Intellectuals}}'}.
\newblock In \bibinfo{booktitle}{\emph{French {{Intellectuals}} and
  {{Politics}} from the {{Dreyfus Affair}} to the {{Occupation}}}},
  \bibfield{editor}{\bibinfo{person}{David Drake}} (Ed.).
  \bibinfo{publisher}{Palgrave Macmillan UK}, \bibinfo{address}{London},
  \bibinfo{pages}{8--34}.
\newblock
\showISBNx{978-0-230-00609-6}
\urldef\tempurl%
\url{https://doi.org/10.1057/9780230006096_2}
\showDOI{\tempurl}


\bibitem[Du and Bozeman(2025)]%
        {DuBozeman2025}
\bibfield{author}{\bibinfo{person}{Luyu Du} {and} \bibinfo{person}{Barry
  Bozeman}.} \bibinfo{year}{2025}\natexlab{}.
\newblock \showarticletitle{Trump's Destruction of {{U}}.{{S}}. Universities:
  Domestic and International Implications}.
\newblock \bibinfo{journal}{\emph{Global Public Policy and Governance}}
  \bibinfo{volume}{5}, \bibinfo{number}{4} (\bibinfo{date}{Dec.}
  \bibinfo{year}{2025}), \bibinfo{pages}{380--392}.
\newblock
\showISSN{2730-6291, 2730-6305}
\urldef\tempurl%
\url{https://doi.org/10.1007/s43508-025-00136-6}
\showDOI{\tempurl}


\bibitem[Edwards(2021)]%
        {Edwards2021}
\bibfield{author}{\bibinfo{person}{Jane Edwards}.}
  \bibinfo{year}{2021}\natexlab{}.
\newblock \showarticletitle{Ethical {{Autoethnography}}: {{Is}} It
  {{Possible}}?}
\newblock \bibinfo{journal}{\emph{International Journal of Qualitative
  Methods}}  \bibinfo{volume}{20} (\bibinfo{date}{Jan.} \bibinfo{year}{2021}),
  \bibinfo{pages}{160940692199530}.
\newblock
\showISSN{1609-4069, 1609-4069}
\urldef\tempurl%
\url{https://doi.org/10.1177/1609406921995306}
\showDOI{\tempurl}


\bibitem[Ellis(2003)]%
        {Ellis2003}
\bibfield{author}{\bibinfo{person}{Carolyn Ellis}.}
  \bibinfo{year}{2003}\natexlab{}.
\newblock \bibinfo{booktitle}{\emph{The {{Ethnographic I}}: {{A Methodological
  Novel}} about {{Autoethnography}}}}.
\newblock \bibinfo{publisher}{AltaMira Press}, \bibinfo{address}{Lanham, MD,
  USA}.
\newblock
\showISBNx{978-0-7591-0051-0}
\urldef\tempurl%
\url{https://rowman.com/ISBN/9780759115866}
\showURL{%
\tempurl}


\bibitem[{EPIC}(2020)]%
        {EPIC2020}
\bibfield{author}{\bibinfo{person}{{EPIC}}.} \bibinfo{year}{2020}\natexlab{}.
\newblock \bibinfo{title}{Liberty at {{Risk}}: {{Pre-trial Risk Assessment
  Tools}} in the {{U}}.{{S}}.}
\newblock
\newblock
\urldef\tempurl%
\url{https://epic.org/documents/liberty-at-risk/}
\showURL{%
\tempurl}


\bibitem[Ernsthausen et~al\mbox{.}(2026)]%
        {Sandoval2026}
\bibfield{author}{\bibinfo{person}{Jeff Ernsthausen}, \bibinfo{person}{Mario
  Ariza}, \bibinfo{person}{McKenzie Funk}, \bibinfo{person}{Mica Rosenberg},
  {and} \bibinfo{person}{Gabriel Sandoval}.} \bibinfo{year}{2026}\natexlab{}.
\newblock \bibinfo{title}{Trump {{Has Detained}} the {{Parents}} of {{More
  Than}} 11,000 {{U}}.{{S}}. {{Citizen Kids}}}.
\newblock
\newblock
\urldef\tempurl%
\url{https://www.propublica.org/article/trump-family-deportations-ice-citizen-kids}
\showURL{%
\tempurl}


\bibitem[Eubanks(2018)]%
        {Eubanks2018}
\bibfield{author}{\bibinfo{person}{Virginia Eubanks}.}
  \bibinfo{year}{2018}\natexlab{}.
\newblock \bibinfo{booktitle}{\emph{Automating {{Inequality}}: {{How High-Tech
  Tools Profile}}, {{Police}}, and {{Punish}} the {{Poor}}}}.
\newblock \bibinfo{publisher}{St. Martin's Press}, \bibinfo{address}{New York,
  NY, USA}.
\newblock
\showISBNx{978-1-250-07431-7}
\urldef\tempurl%
\url{https://us.macmillan.com/books/9781250215789/automating-inequality}
\showURL{%
\tempurl}


\bibitem[Foucault(1979)]%
        {Foucault1979}
\bibfield{author}{\bibinfo{person}{Michel Foucault}.}
  \bibinfo{year}{1979}\natexlab{}.
\newblock \bibinfo{booktitle}{\emph{Discipline and {{Punish}}}}.
\newblock \bibinfo{publisher}{Vintage Books}, \bibinfo{address}{New York, NY,
  USA}.
\newblock
\showISBNx{0-394-72767-3}


\bibitem[Foy and Maney(2025)]%
        {Maney2025}
\bibfield{author}{\bibinfo{person}{Nicole Foy} {and} \bibinfo{person}{Sarabeth
  Maney}.} \bibinfo{year}{2025}\natexlab{}.
\newblock \bibinfo{title}{More {{Than}} 170 {{U}}.{{S}}. {{Citizens Have Been
  Held}} by {{Immigration Agents}}. {{They}}'ve {{Been Kicked}}, {{Dragged}}
  and {{Detained}} for {{Days}}.}
\newblock
\newblock
\urldef\tempurl%
\url{https://www.propublica.org/article/immigration-dhs-american-citizens-arrested-detained-against-will}
\showURL{%
\tempurl}


\bibitem[Francis~Ward(2021)]%
        {FrancisWard2021b}
\bibfield{author}{\bibinfo{person}{Stephanie Francis~Ward}.}
  \bibinfo{year}{2021}\natexlab{}.
\newblock \showarticletitle{While Many Jurisdictions Had Few or No Online Bar
  Exam Testing Violations, {{California}} Had Many}.
\newblock \bibinfo{journal}{\emph{ABA Journal}} (\bibinfo{date}{Jan.}
  \bibinfo{year}{2021}).
\newblock
\urldef\tempurl%
\url{https://www.abajournal.com/web/article/while-many-jurisdictions-had-few-or-no-online-bar-exam-testing-violations-california-had-many}
\showURL{%
\tempurl}


\bibitem[Furlong(1913)]%
        {Furlong1913}
\bibfield{author}{\bibinfo{person}{Charles~Wellington Furlong}.}
  \bibinfo{year}{1913}\natexlab{}.
\newblock \showarticletitle{Cayenne --- the Dry Guillotine}.
\newblock \bibinfo{journal}{\emph{Harper's Magazine}} \bibinfo{volume}{127},
  \bibinfo{number}{757} (\bibinfo{date}{June} \bibinfo{year}{1913}),
  \bibinfo{pages}{3--16}.
\newblock
\urldef\tempurl%
\url{https://archive.org/details/harpersnew127various/page/3/mode/1up}
\showURL{%
\tempurl}


\bibitem[Gebru et~al\mbox{.}(2021)]%
        {GebruMorgensternea2021}
\bibfield{author}{\bibinfo{person}{Timnit Gebru}, \bibinfo{person}{Jamie
  Morgenstern}, \bibinfo{person}{Briana Vecchione},
  \bibinfo{person}{Jennifer~Wortman Vaughan}, \bibinfo{person}{Hanna Wallach},
  \bibinfo{person}{Hal Daum{\'e}~III}, {and} \bibinfo{person}{Kate Crawford}.}
  \bibinfo{year}{2021}\natexlab{}.
\newblock \bibinfo{title}{Datasheets for {{Datasets}}}.
\newblock
\newblock
\urldef\tempurl%
\url{https://doi.org/10.48550/arXiv.1803.09010}
\showDOI{\tempurl}
\showeprint[arxiv]{1803.09010}~[cs]


\bibitem[Gillespie(2018)]%
        {Gillespie2018}
\bibfield{author}{\bibinfo{person}{Tarleton Gillespie}.}
  \bibinfo{year}{2018}\natexlab{}.
\newblock \bibinfo{booktitle}{\emph{Custodians of the {{Internet}}:
  {{Platforms}}, {{Content Moderation}}, and the {{Hidden Decisions That Shape
  Social Media}}}}.
\newblock \bibinfo{publisher}{Yale University Press}, \bibinfo{address}{New
  Haven, CT, USA}.
\newblock
\showISBNx{978-0-300-23502-9}


\bibitem[{Google}(2022)]%
        {Google2022}
\bibfield{author}{\bibinfo{person}{{Google}}.} \bibinfo{year}{2022}\natexlab{}.
\newblock \bibinfo{booktitle}{\emph{{{YouTube Community Guidelines
  Enforcement}}}}.
\newblock \bibinfo{type}{Transparency {{Report}}}.
  \bibinfo{institution}{Google}.
\newblock
\urldef\tempurl%
\url{https://transparencyreport.google.com/youtube-policy/removals?hl=en}
\showURL{%
\tempurl}


\bibitem[Grauer(2021)]%
        {Grauer2021}
\bibfield{author}{\bibinfo{person}{Yael Grauer}.}
  \bibinfo{year}{2021}\natexlab{}.
\newblock \showarticletitle{Millions of {{Leaked Police Files Detail
  Suffocating Surveillance}} of {{China}}'s {{Uyghur Minority}}}.
\newblock \bibinfo{journal}{\emph{The Intercept}} (\bibinfo{date}{Jan.}
  \bibinfo{year}{2021}).
\newblock
\urldef\tempurl%
\url{https://theintercept.com/2021/01/29/china-uyghur-muslim-surveillance-police/}
\showURL{%
\tempurl}


\bibitem[Green et~al\mbox{.}(2022)]%
        {GreenProcopeea2022}
\bibfield{author}{\bibinfo{person}{Nekesha Green}, \bibinfo{person}{Chavez
  Procope}, \bibinfo{person}{Adeel Cheema}, {and} \bibinfo{person}{Adediji
  Adekunle}.} \bibinfo{year}{2022}\natexlab{}.
\newblock \bibinfo{title}{System {{Cards}}, a New Resource for Understanding
  How {{AI}} Systems Work}.
\newblock
\newblock
\urldef\tempurl%
\url{https://ai.facebook.com/blog/system-cards-a-new-resource-for-understanding-how-ai-systems-work/}
\showURL{%
\tempurl}


\bibitem[Hall(2012)]%
        {Hall2012}
\bibfield{author}{\bibinfo{person}{Patrick Hall}.}
  \bibinfo{year}{2012}\natexlab{}.
\newblock \showarticletitle{Incident 135: {{University}} of {{Texas}} at
  {{Austin}}'s {{Algorithm}} to {{Evaluate Graduate Applications}}, {{GRADE}},
  {{Allegedly Exacerbated Existing Inequality}} for {{Marginalized
  Applicants}}, {{Prompting Tool Suspension}}}.
\newblock \bibinfo{journal}{\emph{Artificial Intelligence Incident Database}}
  (\bibinfo{date}{Dec.} \bibinfo{year}{2012}).
\newblock
\urldef\tempurl%
\url{https://incidentdatabase.ai/cite/135}
\showURL{%
\tempurl}


\bibitem[Hall(2020)]%
        {Hall2020a}
\bibfield{author}{\bibinfo{person}{Patrick Hall}.}
  \bibinfo{year}{2020}\natexlab{}.
\newblock \showarticletitle{Incident 86: {{Coding Errors}} in {{Leaving
  Certificate Grading Algorithm Caused Inaccurate Scores}} in {{Ireland}}}.
\newblock \bibinfo{journal}{\emph{Artificial Intelligence Incident Database}}
  (\bibinfo{date}{Oct.} \bibinfo{year}{2020}).
\newblock
\urldef\tempurl%
\url{https://incidentdatabase.ai/cite/86}
\showURL{%
\tempurl}


\bibitem[Hao(2019)]%
        {Hao2019}
\bibfield{author}{\bibinfo{person}{Karen Hao}.}
  \bibinfo{year}{2019}\natexlab{}.
\newblock \showarticletitle{{{AI}} Is Sending People to Jail---and Getting It
  Wrong}.
\newblock \bibinfo{journal}{\emph{MIT Technology Review}} (\bibinfo{date}{Jan.}
  \bibinfo{year}{2019}).
\newblock
\urldef\tempurl%
\url{https://www.technologyreview.com/2019/01/21/137783/algorithms-criminal-justice-ai/}
\showURL{%
\tempurl}


\bibitem[Hao(2020)]%
        {Hao2020}
\bibfield{author}{\bibinfo{person}{Karen Hao}.}
  \bibinfo{year}{2020}\natexlab{}.
\newblock \showarticletitle{The Messy, Secretive Reality behind {{OpenAI}}'s
  Bid to Save the World}.
\newblock \bibinfo{journal}{\emph{MIT Technology Review}} (\bibinfo{date}{Feb.}
  \bibinfo{year}{2020}).
\newblock
\urldef\tempurl%
\url{https://www.technologyreview.com/2020/02/17/844721/ai-openai-moonshot-elon-musk-sam-altman-greg-brockman-messy-secretive-reality/}
\showURL{%
\tempurl}


\bibitem[Hao and Freischlad(2022)]%
        {HaoFreischlad2022}
\bibfield{author}{\bibinfo{person}{Karen Hao} {and} \bibinfo{person}{Nadine
  Freischlad}.} \bibinfo{year}{2022}\natexlab{}.
\newblock \showarticletitle{The Gig Workers Fighting Back against the
  Algorithms}.
\newblock \bibinfo{journal}{\emph{MIT Technology Review}}
  (\bibinfo{date}{April} \bibinfo{year}{2022}).
\newblock
\urldef\tempurl%
\url{https://www.technologyreview.com/2022/04/21/1050381/the-gig-workers-fighting-back-against-the-algorithms/}
\showURL{%
\tempurl}


\bibitem[Hao and Hern{\'a}ndez(2022)]%
        {HaoHernandez2022}
\bibfield{author}{\bibinfo{person}{Karen Hao} {and}
  \bibinfo{person}{Andrea~Paola Hern{\'a}ndez}.}
  \bibinfo{year}{2022}\natexlab{}.
\newblock \showarticletitle{How the {{AI}} Industry Profits from Catastrophe}.
\newblock \bibinfo{journal}{\emph{MIT Technology Review}}
  (\bibinfo{date}{April} \bibinfo{year}{2022}).
\newblock
\urldef\tempurl%
\url{https://www.technologyreview.com/2022/04/20/1050392/ai-industry-appen-scale-data-labels/}
\showURL{%
\tempurl}


\bibitem[Hao and Swart(2022)]%
        {HaoSwart2022}
\bibfield{author}{\bibinfo{person}{Karen Hao} {and} \bibinfo{person}{Heidi
  Swart}.} \bibinfo{year}{2022}\natexlab{}.
\newblock \showarticletitle{South {{Africa}}'s Private Surveillance Machine Is
  Fueling a Digital Apartheid}.
\newblock \bibinfo{journal}{\emph{MIT Technology Review}}
  (\bibinfo{date}{April} \bibinfo{year}{2022}).
\newblock
\urldef\tempurl%
\url{https://www.technologyreview.com/2022/04/19/1049996/south-africa-ai-surveillance-digital-apartheid/}
\showURL{%
\tempurl}


\bibitem[Interlandi(2026)]%
        {Interlandi2026}
\bibfield{author}{\bibinfo{person}{Jeneen Interlandi}.}
  \bibinfo{year}{2026}\natexlab{}.
\newblock \showarticletitle{`{{A Mass Disaster Nonstop}}': {{Inside}} the
  {{Turmoil}} at {{Robert F}}. {{Kennedy Jr}}.'s {{C}}.{{D}}.{{C}}.}
\newblock \bibinfo{journal}{\emph{The New York Times}} (\bibinfo{date}{March}
  \bibinfo{year}{2026}).
\newblock
\showISSN{0362-4331}
\urldef\tempurl%
\url{https://www.nytimes.com/interactive/2026/03/23/magazine/trump-rfk-jr-cdc-vaccines-maha.html}
\showURL{%
\tempurl}


\bibitem[{itsmevic111}(2024)]%
        {itsmevic1112024}
\bibfield{author}{\bibinfo{person}{{itsmevic111}}.}
  \bibinfo{year}{2024}\natexlab{}.
\newblock \bibinfo{title}{Pins Removed for Supposed "Sexual Content"?}
\newblock
\newblock
\urldef\tempurl%
\url{https://web.archive.org/web/20241006041806/https://www.reddit.com/r/Pinterest/comments/1b81v0c/pins_removed_for_supposed_sexual_content/}
\showURL{%
\tempurl}


\bibitem[{Jacedayton}(2024)]%
        {Jacedayton2024}
\bibfield{author}{\bibinfo{person}{{Jacedayton}}.}
  \bibinfo{year}{2024}\natexlab{}.
\newblock \bibinfo{title}{Nudity Policy Re. Art}.
\newblock
\newblock
\urldef\tempurl%
\url{https://web.archive.org/web/20250222235903/https://www.reddit.com/r/Pinterest/comments/1baqmha/nudity_policy_re_art/}
\showURL{%
\tempurl}


\bibitem[Kafka(1995)]%
        {Kafka1995}
\bibfield{author}{\bibinfo{person}{Franz Kafka}.}
  \bibinfo{year}{1995}\natexlab{}.
\newblock \showarticletitle{In the {{Penal Colony}}}.
\newblock In \bibinfo{booktitle}{\emph{The {{Complete Stories}}}}.
  \bibinfo{publisher}{Knopf Doubleday Publishing Group}, \bibinfo{address}{New
  York, NY, USA}.
\newblock
\showISBNx{978-0-8052-1055-2}
\urldef\tempurl%
\url{https://www.kafka-online.info/in-the-penal-colony.html}
\showURL{%
\tempurl}


\bibitem[Kantor et~al\mbox{.}(2022)]%
        {KantorSundaramea2022}
\bibfield{author}{\bibinfo{person}{Jodi Kantor}, \bibinfo{person}{Arya
  Sundaram}, \bibinfo{person}{Aliza Aufrichtig}, {and} \bibinfo{person}{Rumsey
  Taylor}.} \bibinfo{year}{2022}\natexlab{}.
\newblock \showarticletitle{The {{Rise}} of the {{Worker Productivity Score}}}.
\newblock \bibinfo{journal}{\emph{The New York Times}} (\bibinfo{date}{Aug.}
  \bibinfo{year}{2022}).
\newblock
\showISSN{0362-4331}
\urldef\tempurl%
\url{https://www.nytimes.com/interactive/2022/08/14/business/worker-productivity-tracking.html}
\showURL{%
\tempurl}


\bibitem[Kantor et~al\mbox{.}(2021)]%
        {KantorWeiseea2021}
\bibfield{author}{\bibinfo{person}{Jodi Kantor}, \bibinfo{person}{Karen Weise},
  {and} \bibinfo{person}{Grace Ashford}.} \bibinfo{year}{2021}\natexlab{}.
\newblock \showarticletitle{The {{Amazon That Customers Don}}'t {{See}}}.
\newblock \bibinfo{journal}{\emph{The New York Times}} (\bibinfo{date}{June}
  \bibinfo{year}{2021}).
\newblock
\showISSN{0362-4331}
\urldef\tempurl%
\url{https://www.nytimes.com/interactive/2021/06/15/us/amazon-workers.html}
\showURL{%
\tempurl}


\bibitem[Katsaros et~al\mbox{.}(2022)]%
        {KatsarosTylerea2022}
\bibfield{author}{\bibinfo{person}{Matthew Katsaros}, \bibinfo{person}{Tom
  Tyler}, \bibinfo{person}{Jisu Kim}, {and} \bibinfo{person}{Tracey Meares}.}
  \bibinfo{year}{2022}\natexlab{}.
\newblock \showarticletitle{Procedural {{Justice}} and {{Self Governance}} on
  {{Twitter}}: {{Unpacking}} the {{Experience}} of {{Rule Breaking}} on
  {{Twitter}}}.
\newblock \bibinfo{journal}{\emph{Journal of Online Trust and Safety}}
  \bibinfo{volume}{1}, \bibinfo{number}{3} (\bibinfo{date}{Aug.}
  \bibinfo{year}{2022}).
\newblock
\showISSN{2770-3142}
\urldef\tempurl%
\url{https://doi.org/10.54501/jots.v1i3.38}
\showDOI{\tempurl}


\bibitem[Kellerhoff(2022)]%
        {Kellerhoff2022}
\bibfield{author}{\bibinfo{person}{Sven~Felix Kellerhoff}.}
  \bibinfo{year}{2022}\natexlab{}.
\newblock \showarticletitle{{Spitzelstaat DDR: Die Stasi hatte viel mehr
  Informanten als bekannt}}.
\newblock \bibinfo{journal}{\emph{Die Welt}} (\bibinfo{date}{Jan.}
  \bibinfo{year}{2022}).
\newblock
\urldef\tempurl%
\url{https://www.welt.de/geschichte/article132502154/Spitzelstaat-DDR-Die-Stasi-hatte-viel-mehr-Informanten-als-bekannt.html}
\showURL{%
\tempurl}


\bibitem[Keyes et~al\mbox{.}(2019)]%
        {KeyesHutsonea2019}
\bibfield{author}{\bibinfo{person}{Os Keyes}, \bibinfo{person}{Jevan Hutson},
  {and} \bibinfo{person}{Meredith Durbin}.} \bibinfo{year}{2019}\natexlab{}.
\newblock \showarticletitle{A {{Mulching Proposal}}: {{Analysing}} and
  {{Improving}} an {{Algorithmic System}} for {{Turning}} the {{Elderly}} into
  {{High-Nutrient Slurry}}}. In \bibinfo{booktitle}{\emph{Extended
  {{Abstracts}} of the 2019 {{CHI Conference}} on {{Human Factors}} in
  {{Computing Systems}}}} \emph{(\bibinfo{series}{{{CHI EA}} '19})}.
  \bibinfo{publisher}{Association for Computing Machinery},
  \bibinfo{address}{New York, NY, USA}, \bibinfo{pages}{1--11}.
\newblock
\showISBNx{978-1-4503-5971-9}
\urldef\tempurl%
\url{https://doi.org/10.1145/3290607.3310433}
\showDOI{\tempurl}


\bibitem[Klonick(2018)]%
        {Klonick2018}
\bibfield{author}{\bibinfo{person}{Kate Klonick}.}
  \bibinfo{year}{2018}\natexlab{}.
\newblock \showarticletitle{The {{New Governors}}: {{The People}}, {{Rules}},
  and {{Processes Governing Online Speech}}}.
\newblock \bibinfo{journal}{\emph{Harvard Law Review}} \bibinfo{volume}{131},
  \bibinfo{number}{6} (\bibinfo{date}{April} \bibinfo{year}{2018}),
  \bibinfo{pages}{1598--1670}.
\newblock
\urldef\tempurl%
\url{https://harvardlawreview.org/2018/04/the-new-governors-the-people-rules-and-processes-governing-online-speech/}
\showURL{%
\tempurl}


\bibitem[Kovarsky(2022)]%
        {Kovarsky2022}
\bibfield{author}{\bibinfo{person}{Lee Kovarsky}.}
  \bibinfo{year}{2022}\natexlab{}.
\newblock \showarticletitle{The {{Trump Executions}}}.
\newblock \bibinfo{journal}{\emph{Texas Law Review}} \bibinfo{volume}{100},
  \bibinfo{number}{4} (\bibinfo{date}{March} \bibinfo{year}{2022}).
\newblock
\urldef\tempurl%
\url{https://texaslawreview.org/the-trump-executions/}
\showURL{%
\tempurl}


\bibitem[Kugel(2022)]%
        {Kugel2022}
\bibfield{author}{\bibinfo{person}{Seth Kugel}.}
  \bibinfo{year}{2022}\natexlab{}.
\newblock \showarticletitle{Help! {{I Was Banned From Lyft}} and {{No One Will
  Tell Me Why}}.}
\newblock \bibinfo{journal}{\emph{The New York Times}} (\bibinfo{date}{Nov.}
  \bibinfo{year}{2022}).
\newblock
\showISSN{0362-4331}
\urldef\tempurl%
\url{https://www.nytimes.com/2022/11/17/travel/tripped-up-banned-lyft.html}
\showURL{%
\tempurl}


\bibitem[Lam(2020)]%
        {Lam2020}
\bibfield{author}{\bibinfo{person}{Khoa Lam}.} \bibinfo{year}{2020}\natexlab{}.
\newblock \showarticletitle{Incident 374: {{UK Ofqual}}'s {{Algorithm
  Disproportionately Provided Lower Grades Than Teachers}}' {{Assessments}}}.
\newblock \bibinfo{journal}{\emph{Artificial Intelligence Incident Database}}
  (\bibinfo{date}{Aug.} \bibinfo{year}{2020}).
\newblock
\urldef\tempurl%
\url{https://incidentdatabase.ai/cite/374}
\showURL{%
\tempurl}


\bibitem[Lea(1906a)]%
        {Lea1906a}
\bibfield{author}{\bibinfo{person}{Henry~Charles Lea}.}
  \bibinfo{year}{1906}\natexlab{a}.
\newblock \bibinfo{booktitle}{\emph{A History of the {{Inquisition}} of
  {{Spain}}}}. Vol.~\bibinfo{volume}{1}.
\newblock \bibinfo{publisher}{The Macmillan Company}, \bibinfo{address}{New
  York, NY, USA}.
\newblock
\urldef\tempurl%
\url{http://archive.org/details/ahistoryinquisi03leagoog}
\showURL{%
\tempurl}


\bibitem[Lea(1906b)]%
        {Lea1906b}
\bibfield{author}{\bibinfo{person}{Henry~Charles Lea}.}
  \bibinfo{year}{1906}\natexlab{b}.
\newblock \bibinfo{booktitle}{\emph{A History of the {{Inquisition}} of
  {{Spain}}}}. Vol.~\bibinfo{volume}{2}.
\newblock \bibinfo{publisher}{The Macmillan Company}, \bibinfo{address}{New
  York, NY, USA}.
\newblock
\urldef\tempurl%
\url{http://archive.org/details/historyofinquisispain02leah}
\showURL{%
\tempurl}


\bibitem[Lea(1906c)]%
        {Lea1906c}
\bibfield{author}{\bibinfo{person}{Henry~Charles Lea}.}
  \bibinfo{year}{1906}\natexlab{c}.
\newblock \bibinfo{booktitle}{\emph{A History of the {{Inquisition}} of
  {{Spain}}}}. Vol.~\bibinfo{volume}{4}.
\newblock \bibinfo{publisher}{The Macmillan Company}, \bibinfo{address}{New
  York, NY, USA}.
\newblock
\showLCCN{1185366}
\urldef\tempurl%
\url{http://archive.org/details/historyofinquisi04leah}
\showURL{%
\tempurl}


\bibitem[Lea(1906d)]%
        {Lea1906d}
\bibfield{author}{\bibinfo{person}{Henry~Charles Lea}.}
  \bibinfo{year}{1906}\natexlab{d}.
\newblock \bibinfo{booktitle}{\emph{A History of the {{Inquisition}} of
  {{Spain}}}}. Vol.~\bibinfo{volume}{3}.
\newblock \bibinfo{publisher}{The Macmillan Company}, \bibinfo{address}{New
  York, NY, USA}.
\newblock
\showLCCN{SRLF:LAGE-1948466}
\urldef\tempurl%
\url{http://archive.org/details/historyofinquisi03leahiala}
\showURL{%
\tempurl}


\bibitem[Lennard(2020)]%
        {Lennard2020}
\bibfield{author}{\bibinfo{person}{Natasha Lennard}.}
  \bibinfo{year}{2020}\natexlab{}.
\newblock \showarticletitle{Amazon {{Workers Are Organizing}} a {{Global
  Struggle}}}.
\newblock \bibinfo{journal}{\emph{The Intercept}} (\bibinfo{date}{Dec.}
  \bibinfo{year}{2020}).
\newblock
\urldef\tempurl%
\url{https://theintercept.com/2020/12/03/amazon-workers-union-international-strike/}
\showURL{%
\tempurl}


\bibitem[Mackey(2025)]%
        {Mackey2025}
\bibfield{author}{\bibinfo{person}{Robert Mackey}.}
  \bibinfo{year}{2025}\natexlab{}.
\newblock \showarticletitle{French Scientist Denied {{US}} Entry after Phone
  Messages Critical of {{Trump}} Found}.
\newblock \bibinfo{journal}{\emph{The Guardian}} (\bibinfo{date}{March}
  \bibinfo{year}{2025}).
\newblock
\showISSN{0261-3077}
\urldef\tempurl%
\url{https://www.theguardian.com/us-news/2025/mar/19/trump-musk-french-scientist-detained}
\showURL{%
\tempurl}


\bibitem[Masnick(2019)]%
        {Masnick2019}
\bibfield{author}{\bibinfo{person}{Mike Masnick}.}
  \bibinfo{year}{2019}\natexlab{}.
\newblock \bibinfo{title}{Masnick's {{Impossibility Theorem}}: {{Content
  Moderation At Scale Is Impossible To Do Well}}}.
\newblock
\newblock
\urldef\tempurl%
\url{https://www.techdirt.com/2019/11/20/masnicks-impossibility-theorem-content-moderation-scale-is-impossible-to-do-well/}
\showURL{%
\tempurl}


\bibitem[McElroy et~al\mbox{.}(2021)]%
        {McElroyWhittakerea2021}
\bibfield{author}{\bibinfo{person}{Erin McElroy}, \bibinfo{person}{Meredith
  Whittaker}, {and} \bibinfo{person}{Nicole~E. Weber}.}
  \bibinfo{year}{2021}\natexlab{}.
\newblock \bibinfo{title}{Prison {{Tech Comes Home}}}.
\newblock
\newblock
\urldef\tempurl%
\url{https://www.publicbooks.org/prison-tech-comes-home/}
\showURL{%
\tempurl}


\bibitem[Mirbeau(2008)]%
        {Mirbeau2008}
\bibfield{author}{\bibinfo{person}{Octave Mirbeau}.}
  \bibinfo{year}{2008}\natexlab{}.
\newblock \bibinfo{booktitle}{\emph{The {{Torture Garden}}}}.
\newblock \bibinfo{publisher}{Bookkake}, \bibinfo{address}{London, United
  Kingdom}.
\newblock
\showISBNx{978-1-906110-02-4}
\urldef\tempurl%
\url{https://www.bauerverlag.eu/downloads/torture-garden.pdf}
\showURL{%
\tempurl}


\bibitem[Mishkin and Ahmad(2022)]%
        {MishkinAhmad2022}
\bibfield{author}{\bibinfo{person}{Pamela Mishkin} {and} \bibinfo{person}{Lama
  Ahmad}.} \bibinfo{year}{2022}\natexlab{}.
\newblock \bibinfo{title}{{{DALL-E}} 2 {{Preview}}: {{Risks}} and
  {{Limitations}}}.
\newblock
\newblock
\urldef\tempurl%
\url{https://github.com/openai/dalle-2-preview/blob/eeec5a1843b1d17cb9ed113117a2fcaa9206a564/system-card.md}
\showURL{%
\tempurl}


\bibitem[Mitchell et~al\mbox{.}(2019)]%
        {MitchellWuea2019}
\bibfield{author}{\bibinfo{person}{Margaret Mitchell}, \bibinfo{person}{Simone
  Wu}, \bibinfo{person}{Andrew Zaldivar}, \bibinfo{person}{Parker Barnes},
  \bibinfo{person}{Lucy Vasserman}, \bibinfo{person}{Ben Hutchinson},
  \bibinfo{person}{Elena Spitzer}, \bibinfo{person}{Inioluwa~Deborah Raji},
  {and} \bibinfo{person}{Timnit Gebru}.} \bibinfo{year}{2019}\natexlab{}.
\newblock \showarticletitle{Model {{Cards}} for {{Model Reporting}}}. In
  \bibinfo{booktitle}{\emph{Proceedings of the {{Conference}} on {{Fairness}},
  {{Accountability}}, and {{Transparency}}}} \emph{(\bibinfo{series}{{{FAccT}}
  '19})}. \bibinfo{publisher}{Association for Computing Machinery},
  \bibinfo{address}{New York, NY, USA}, \bibinfo{pages}{220--229}.
\newblock
\showISBNx{978-1-4503-6125-5}
\urldef\tempurl%
\url{https://doi.org/10.1145/3287560.3287596}
\showDOI{\tempurl}


\bibitem[Mori et~al\mbox{.}(2023)]%
        {MoriParkea2023}
\bibfield{author}{\bibinfo{person}{Camille Mori}, \bibinfo{person}{Julianna
  Park}, \bibinfo{person}{Nicole Racine}, \bibinfo{person}{Heather Ganshorn},
  \bibinfo{person}{Cailey Hartwick}, {and} \bibinfo{person}{Sheri Madigan}.}
  \bibinfo{year}{2023}\natexlab{}.
\newblock \showarticletitle{Exposure to Sexual Content and Problematic Sexual
  Behaviors in Children and Adolescents: {{A}} Systematic Review and
  Meta-Analysis}.
\newblock \bibinfo{journal}{\emph{Child Abuse \& Neglect}}
  \bibinfo{volume}{143} (\bibinfo{date}{Sept.} \bibinfo{year}{2023}),
  \bibinfo{pages}{106255}.
\newblock
\showISSN{01452134}
\urldef\tempurl%
\url{https://doi.org/10.1016/j.chiabu.2023.106255}
\showDOI{\tempurl}


\bibitem[Mozur et~al\mbox{.}(2022)]%
        {MozurXiaoea2022}
\bibfield{author}{\bibinfo{person}{Paul Mozur}, \bibinfo{person}{Muyi Xiao},
  {and} \bibinfo{person}{John Liu}.} \bibinfo{year}{2022}\natexlab{}.
\newblock \showarticletitle{`{{An Invisible Cage}}': {{How China Is Policing}}
  the {{Future}}}.
\newblock \bibinfo{journal}{\emph{The New York Times}} (\bibinfo{date}{June}
  \bibinfo{year}{2022}).
\newblock
\showISSN{0362-4331}
\urldef\tempurl%
\url{https://www.nytimes.com/2022/06/25/technology/china-surveillance-police.html}
\showURL{%
\tempurl}


\bibitem[Myers(2025)]%
        {Myers2025}
\bibfield{author}{\bibinfo{person}{Steven~Lee Myers}.}
  \bibinfo{year}{2025}\natexlab{}.
\newblock \showarticletitle{U.{{S}}. {{Bars}} 5 {{European Tech Regulators}}
  and {{Researchers}}}.
\newblock \bibinfo{journal}{\emph{The New York Times}} (\bibinfo{date}{Dec.}
  \bibinfo{year}{2025}).
\newblock
\showISSN{0362-4331}
\urldef\tempurl%
\url{https://www.nytimes.com/2025/12/23/technology/trump-rubio-european-tech-disinformation-digital-services-act.html}
\showURL{%
\tempurl}


\bibitem[Nagy and Neff(2024)]%
        {NagyNeff2024}
\bibfield{author}{\bibinfo{person}{Peter Nagy} {and} \bibinfo{person}{Gina
  Neff}.} \bibinfo{year}{2024}\natexlab{}.
\newblock \showarticletitle{Conjuring Algorithms: {{Understanding}} the Tech
  Industry as Stage Magicians}.
\newblock \bibinfo{journal}{\emph{New Media \& Society}} \bibinfo{volume}{26},
  \bibinfo{number}{9} (\bibinfo{date}{Sept.} \bibinfo{year}{2024}),
  \bibinfo{pages}{4938--4954}.
\newblock
\showISSN{1461-4448, 1461-7315}
\urldef\tempurl%
\url{https://doi.org/10.1177/14614448241251789}
\showDOI{\tempurl}


\bibitem[Narayanan and Kapoor(2024)]%
        {NarayananKapoor2024}
\bibfield{author}{\bibinfo{person}{Arvind Narayanan} {and}
  \bibinfo{person}{Sayash Kapoor}.} \bibinfo{year}{2024}\natexlab{}.
\newblock \bibinfo{booktitle}{\emph{{{AI Snake Oil}}: {{What Artificial
  Intelligence Can Do}}, {{What It Can}}'t, and {{How}} to {{Tell}} the
  {{Difference}}}}.
\newblock \bibinfo{publisher}{Princeton University Press},
  \bibinfo{address}{Princeton, NJ, USA}.
\newblock
\urldef\tempurl%
\url{https://press.princeton.edu/books/hardcover/9780691249131/ai-snake-oil}
\showURL{%
\tempurl}


\bibitem[{Natalie}(2022)]%
        {Natalie2022}
\bibfield{author}{\bibinfo{person}{{Natalie}}.}
  \bibinfo{year}{2022}\natexlab{}.
\newblock \bibinfo{title}{I Received a Warning While Using
  {{DALL}}{$\cdot$}{{E}} 2. {{Will I}} Be Banned?}
\newblock
\newblock
\urldef\tempurl%
\url{https://web.archive.org/web/20220818131242/https://help.openai.com/en/articles/6338765-i-received-a-warning-while-using-dall-e-2-will-i-be-banned}
\showURL{%
\tempurl}


\bibitem[Olmstead(2022)]%
        {Olmstead2022}
\bibfield{author}{\bibinfo{person}{Molly Olmstead}.}
  \bibinfo{year}{2022}\natexlab{}.
\newblock \showarticletitle{Why {{Did}} 11 {{Billion Alaskan Snow Crabs
  Suddenly Disappear}}?}
\newblock \bibinfo{journal}{\emph{Slate}} (\bibinfo{date}{Oct.}
  \bibinfo{year}{2022}).
\newblock
\showISSN{1091-2339}
\urldef\tempurl%
\url{https://slate.com/technology/2022/10/alaskan-snow-crabs-dead.html}
\showURL{%
\tempurl}


\bibitem[{OpenAI}(2018)]%
        {OpenAI2018}
\bibfield{author}{\bibinfo{person}{{OpenAI}}.} \bibinfo{year}{2018}\natexlab{}.
\newblock \bibinfo{title}{{{OpenAI Charter}}}.
\newblock
\newblock
\urldef\tempurl%
\url{https://openai.com/charter/}
\showURL{%
\tempurl}


\bibitem[{OpenAI}(2022a)]%
        {OpenAI2022h}
\bibfield{author}{\bibinfo{person}{{OpenAI}}.}
  \bibinfo{year}{2022}\natexlab{a}.
\newblock \bibinfo{title}{{{DALL}}{$\cdot$}{{E API Now Available}} in {{Public
  Beta}}}.
\newblock
\newblock
\urldef\tempurl%
\url{https://openai.com/blog/dall-e-api-now-available-in-public-beta/}
\showURL{%
\tempurl}


\bibitem[{OpenAI}(2022b)]%
        {OpenAI2022a}
\bibfield{author}{\bibinfo{person}{{OpenAI}}.}
  \bibinfo{year}{2022}\natexlab{b}.
\newblock \bibinfo{title}{{{DALL}}{$\cdot$}{{E Now Available Without
  Waitlist}}}.
\newblock
\newblock
\urldef\tempurl%
\url{https://openai.com/blog/dall-e-now-available-without-waitlist/}
\showURL{%
\tempurl}


\bibitem[Ortutay(2020)]%
        {Ortutay2020}
\bibfield{author}{\bibinfo{person}{Barbara Ortutay}.}
  \bibinfo{year}{2020}\natexlab{}.
\newblock \bibinfo{title}{Does the Naked Body Belong on {{Facebook}}? {{It}}'s
  Complicated}.
\newblock
\newblock
\urldef\tempurl%
\url{https://apnews.com/general-news-a5bb16de8aeee3253cc45d265d874299}
\showURL{%
\tempurl}


\bibitem[{Owen-Smith}(2025)]%
        {Owen-Smith2025}
\bibfield{author}{\bibinfo{person}{Jason {Owen-Smith}}.}
  \bibinfo{year}{2025}\natexlab{}.
\newblock \showarticletitle{Why {{Universities Are So Powerless}} in {{Their
  Fight Against Trump}}}.
\newblock \bibinfo{journal}{\emph{The Chronicle of Higher Education}}
  (\bibinfo{date}{July} \bibinfo{year}{2025}).
\newblock
\urldef\tempurl%
\url{https://www.chronicle.com/article/why-universities-are-so-powerless-in-their-fight-against-trump}
\showURL{%
\tempurl}


\bibitem[Perrigo(2023)]%
        {Perrigo2023a}
\bibfield{author}{\bibinfo{person}{Billy Perrigo}.}
  \bibinfo{year}{2023}\natexlab{}.
\newblock \showarticletitle{{{OpenAI Used Kenyan Workers}} on {{Less Than}} \$2
  {{Per Hour}} to {{Make ChatGPT Less Toxic}}}.
\newblock \bibinfo{journal}{\emph{Time}} (\bibinfo{date}{Jan.}
  \bibinfo{year}{2023}).
\newblock
\urldef\tempurl%
\url{https://time.com/6247678/openai-chatgpt-kenya-workers/}
\showURL{%
\tempurl}


\bibitem[Perry et~al\mbox{.}(2013)]%
        {PerryMcInnisea2013}
\bibfield{author}{\bibinfo{person}{Walter~L. Perry}, \bibinfo{person}{Brian
  McInnis}, \bibinfo{person}{Carter~C. Price}, \bibinfo{person}{Susan Smith},
  {and} \bibinfo{person}{John~S. Hollywood}.} \bibinfo{year}{2013}\natexlab{}.
\newblock \bibinfo{booktitle}{\emph{Predictive {{Policing}}: {{The Role}} of
  {{Crime Forecasting}} in {{Law Enforcement Operations}}}}.
\newblock \bibinfo{publisher}{RAND Corporation}, \bibinfo{address}{Santa
  Monica, CA}.
\newblock
\urldef\tempurl%
\url{https://doi.org/10.7249/RR233}
\showDOI{\tempurl}


\bibitem[{Pinterest}(2022)]%
        {Pinterest2022}
\bibfield{author}{\bibinfo{person}{{Pinterest}}.}
  \bibinfo{year}{2022}\natexlab{}.
\newblock \bibinfo{booktitle}{\emph{Q1+{{Q2}} 2022 {{Transparency Report}}}}.
\newblock \bibinfo{type}{Transparency {{Report}}} Q1+Q2 2022.
  \bibinfo{institution}{Pinterest}.
\newblock
\urldef\tempurl%
\url{https://policy.pinterest.com/en/transparency-report}
\showURL{%
\tempurl}


\bibitem[{Pinterest}(2025)]%
        {Pinterest2025}
\bibfield{author}{\bibinfo{person}{{Pinterest}}.}
  \bibinfo{year}{2025}\natexlab{}.
\newblock \bibinfo{title}{January--{{June}} 2024 {{Transparency Report}}}.
\newblock
\newblock
\urldef\tempurl%
\url{https://policy.pinterest.com/en/transparency-report-h1-2024}
\showURL{%
\tempurl}


\bibitem[Prescott et~al\mbox{.}(2018)]%
        {PrescottSargentea2018}
\bibfield{author}{\bibinfo{person}{Anna~T. Prescott}, \bibinfo{person}{James~D.
  Sargent}, {and} \bibinfo{person}{Jay~G. Hull}.}
  \bibinfo{year}{2018}\natexlab{}.
\newblock \showarticletitle{Metaanalysis of the Relationship between Violent
  Video Game Play and Physical Aggression over Time}.
\newblock \bibinfo{journal}{\emph{Proceedings of the National Academy of
  Sciences}} \bibinfo{volume}{115}, \bibinfo{number}{40} (\bibinfo{date}{Oct.}
  \bibinfo{year}{2018}), \bibinfo{pages}{9882--9888}.
\newblock
\urldef\tempurl%
\url{https://doi.org/10.1073/pnas.1611617114}
\showDOI{\tempurl}


\bibitem[Procope et~al\mbox{.}(2022)]%
        {ProcopeCheemaea2022}
\bibfield{author}{\bibinfo{person}{Chavez Procope}, \bibinfo{person}{Adeel
  Cheema}, \bibinfo{person}{David Adkins}, \bibinfo{person}{Bilal Alsallakh},
  \bibinfo{person}{Nekesha Green}, \bibinfo{person}{Emily McReynolds},
  \bibinfo{person}{Grace Pehl}, \bibinfo{person}{Erin Wang}, {and}
  \bibinfo{person}{Polina Zvyagina}.} \bibinfo{year}{2022}\natexlab{}.
\newblock \bibinfo{booktitle}{\emph{System-{{Level Transparency}} of {{Machine
  Learning}}}}.
\newblock \bibinfo{type}{{T}echnical {R}eport}. \bibinfo{institution}{Meta}.
\newblock
\urldef\tempurl%
\url{https://ai.facebook.com/research/publications/system-level-transparency-of-machine-learning}
\showURL{%
\tempurl}


\bibitem[Redden et~al\mbox{.}(2020)]%
        {ReddenODonovanDixea2020}
\bibfield{author}{\bibinfo{person}{James Redden}, \bibinfo{person}{Molly
  O'Donovan~Dix}, {and} \bibinfo{person}{{Criminal Justice Testing {and}
  Evaluation Consortium}}.} \bibinfo{year}{2020}\natexlab{}.
\newblock \bibinfo{title}{Artificial {{Intelligence}} in the {{Criminal Justice
  System}}}.
\newblock
\newblock
\urldef\tempurl%
\url{https://cjtec.org/artificial-intelligence-in-the-criminal-justice-system/}
\showURL{%
\tempurl}


\bibitem[Redfield(2005)]%
        {Redfield2005}
\bibfield{author}{\bibinfo{person}{Peter Redfield}.}
  \bibinfo{year}{2005}\natexlab{}.
\newblock \showarticletitle{Foucault in the {{Tropics}}: {{Displacing}} the
  {{Panopticon}}}.
\newblock In \bibinfo{booktitle}{\emph{Anthropologies of {{Modernity}}:
  {{Foucault}}, {{Governmentality}}, and {{Life Politics}}}},
  \bibfield{editor}{\bibinfo{person}{Jonathan~Xavier Inda}} (Ed.).
  \bibinfo{publisher}{Blackwell Publishing}, \bibinfo{address}{Hoboken, NJ,
  USA}, \bibinfo{pages}{50--79}.
\newblock
\showISBNx{978-1-4051-5302-7}
\urldef\tempurl%
\url{https://redfield.web.unc.edu/wp-content/uploads/sites/9305/2015/08/foucault-tropics.pdf}
\showURL{%
\tempurl}


\bibitem[{Ren\'e Belbenoit}(1938)]%
        {ReneBelbenoit1938}
\bibfield{author}{\bibinfo{person}{{Ren\'e Belbenoit}}.}
  \bibinfo{year}{1938}\natexlab{}.
\newblock \bibinfo{booktitle}{\emph{Dry Guillotine: {{Fifteen}} Years among the
  Living Dead}}.
\newblock \bibinfo{publisher}{Blue Ribbon Books}, \bibinfo{address}{New York,
  NY, USA}.
\newblock
\urldef\tempurl%
\url{http://archive.org/details/Dry_Guillotine}
\showURL{%
\tempurl}


\bibitem[Robertson(2017)]%
        {Robertson2017}
\bibfield{author}{\bibinfo{person}{Ritchie Robertson}.}
  \bibinfo{year}{2017}\natexlab{}.
\newblock \showarticletitle{Kafka's {{Reading}}}.
\newblock In \bibinfo{booktitle}{\emph{Franz {{Kafka}} in {{Context}}}},
  \bibfield{editor}{\bibinfo{person}{Carolin Duttlinger}} (Ed.).
  \bibinfo{publisher}{Cambridge University Press}, \bibinfo{address}{Cambridge,
  England}, \bibinfo{pages}{82--90}.
\newblock
\showISBNx{978-1-107-08549-7}
\urldef\tempurl%
\url{https://doi.org/10.1017/9781316084243.011}
\showDOI{\tempurl}


\bibitem[Rodenhizer and Edwards(2019)]%
        {RodenhizerEdwards2019}
\bibfield{author}{\bibinfo{person}{Kara Anne~E. Rodenhizer} {and}
  \bibinfo{person}{Katie~M. Edwards}.} \bibinfo{year}{2019}\natexlab{}.
\newblock \showarticletitle{The {{Impacts}} of {{Sexual Media Exposure}} on
  {{Adolescent}} and {{Emerging Adults}}' {{Dating}} and {{Sexual Violence
  Attitudes}} and {{Behaviors}}: {{A Critical Review}} of the {{Literature}}}.
\newblock \bibinfo{journal}{\emph{Trauma, Violence, \& Abuse}}
  \bibinfo{volume}{20}, \bibinfo{number}{4} (\bibinfo{date}{Oct.}
  \bibinfo{year}{2019}), \bibinfo{pages}{439--452}.
\newblock
\showISSN{1524-8380, 1552-8324}
\urldef\tempurl%
\url{https://doi.org/10.1177/1524838017717745}
\showDOI{\tempurl}


\bibitem[Rosenblat(2018)]%
        {Rosenblat2018}
\bibfield{author}{\bibinfo{person}{Alex Rosenblat}.}
  \bibinfo{year}{2018}\natexlab{}.
\newblock \bibinfo{booktitle}{\emph{Uberland: {{How Algorithms Are Rewriting}}
  the {{Rules}} of {{Work}}} (\bibinfo{edition}{1} ed.)}.
\newblock \bibinfo{publisher}{University of California Press},
  \bibinfo{address}{Berkeley, CA, USA}.
\newblock
\showISBNx{978-0-520-29857-6}
\showeprint[jstor]{10.1525/j.ctv5cgbm3}
\urldef\tempurl%
\url{https://www.jstor.org/stable/10.1525/j.ctv5cgbm3}
\showURL{%
\tempurl}


\bibitem[Sainato(2021)]%
        {Sainato2021}
\bibfield{author}{\bibinfo{person}{Michael Sainato}.}
  \bibinfo{year}{2021}\natexlab{}.
\newblock \showarticletitle{`{{I}}'m Still in Pain': {{Amazon}} Employees Say
  Climate of Fear Has Led to High Rates of Injuries}.
\newblock \bibinfo{journal}{\emph{The Guardian}} (\bibinfo{date}{Dec.}
  \bibinfo{year}{2021}).
\newblock
\showISSN{0261-3077}
\urldef\tempurl%
\url{https://www.theguardian.com/technology/2021/dec/30/amazon-employees-climate-fear-high-rates-injuries}
\showURL{%
\tempurl}


\bibitem[Sainato(2022)]%
        {Sainato2022}
\bibfield{author}{\bibinfo{person}{Michael Sainato}.}
  \bibinfo{year}{2022}\natexlab{}.
\newblock \showarticletitle{Amazon Could Run out of Workers in {{US}} in Two
  Years, Internal Memo Suggests}.
\newblock \bibinfo{journal}{\emph{The Guardian}} (\bibinfo{date}{June}
  \bibinfo{year}{2022}).
\newblock
\showISSN{0261-3077}
\urldef\tempurl%
\url{https://www.theguardian.com/technology/2022/jun/22/amazon-workers-shortage-leaked-memo-warehouse}
\showURL{%
\tempurl}


\bibitem[Saltzer et~al\mbox{.}(1984)]%
        {SaltzerReedea1984}
\bibfield{author}{\bibinfo{person}{J.~H. Saltzer}, \bibinfo{person}{D.~P.
  Reed}, {and} \bibinfo{person}{D.~D. Clark}.} \bibinfo{year}{1984}\natexlab{}.
\newblock \showarticletitle{End-to-End Arguments in System Design}.
\newblock \bibinfo{journal}{\emph{ACM Trans. Comput. Syst.}}
  \bibinfo{volume}{2}, \bibinfo{number}{4} (\bibinfo{date}{Nov.}
  \bibinfo{year}{1984}), \bibinfo{pages}{277--288}.
\newblock
\showISSN{0734-2071}
\urldef\tempurl%
\url{https://doi.org/10.1145/357401.357402}
\showDOI{\tempurl}


\bibitem[Sawyer and Wagner(2022)]%
        {SawyerWagner2022}
\bibfield{author}{\bibinfo{person}{Wendy Sawyer} {and} \bibinfo{person}{Peter
  Wagner}.} \bibinfo{year}{2022}\natexlab{}.
\newblock \bibinfo{title}{Mass {{Incarceration}}: {{The Whole Pie}} 2022}.
\newblock
\newblock
\urldef\tempurl%
\url{https://www.prisonpolicy.org/reports/pie2022.html}
\showURL{%
\tempurl}


\bibitem[Schiffer et~al\mbox{.}(2023)]%
        {SchifferNewtonea2023}
\bibfield{author}{\bibinfo{person}{Zo{\"e} Schiffer}, \bibinfo{person}{Casey
  Newton}, {and} \bibinfo{person}{Alex Heath}.}
  \bibinfo{year}{2023}\natexlab{}.
\newblock \showarticletitle{Inside {{Elon}}'s `{{Extremely Hardcore}}'
  {{Twitter}}}.
\newblock \bibinfo{journal}{\emph{New York Magazine}} (\bibinfo{date}{Jan.}
  \bibinfo{year}{2023}).
\newblock
\urldef\tempurl%
\url{https://nymag.com/intelligencer/article/elon-musk-twitter-takeover.html}
\showURL{%
\tempurl}


\bibitem[Schneider(2025)]%
        {Schneider2025}
\bibfield{author}{\bibinfo{person}{Matti Schneider}.}
  \bibinfo{year}{2025}\natexlab{}.
\newblock \bibinfo{title}{Pinterest Bans All Nudity in the {{US}}, Removing
  Exceptions for Artwork or Education}.
\newblock
\newblock
\urldef\tempurl%
\url{https://opentermsarchive.org/en/memos/pinterest-bans-all-nudity/}
\showURL{%
\tempurl}


\bibitem[Schroeder(2013)]%
        {Schroeder2013}
\bibfield{author}{\bibinfo{person}{Klaus Schroeder}.}
  \bibinfo{year}{2013}\natexlab{}.
\newblock \bibinfo{booktitle}{\emph{Der {{SED-Staat}}: {{Geschicte}} Und
  {{Strukturen}} Der {{DDR}} 1949-1990} (\bibinfo{edition}{3} ed.)}.
\newblock \bibinfo{publisher}{B\"ohlau Verlag}, \bibinfo{address}{Wien, Austria
  and K\"oln, Germany}.
\newblock
\showISBNx{978-3-412-21109-7}
\urldef\tempurl%
\url{https://www.vandenhoeck-ruprecht-verlage.com/themen-entdecken/geschichte/geschichte-der-neuzeit/42210/der-sed-staat}
\showURL{%
\tempurl}


\bibitem[Schuhmann et~al\mbox{.}(2021)]%
        {SchuhmannVencuea2021}
\bibfield{author}{\bibinfo{person}{Christoph Schuhmann},
  \bibinfo{person}{Richard Vencu}, \bibinfo{person}{Romain Beaumont},
  \bibinfo{person}{Robert Kaczmarczyk}, \bibinfo{person}{Clayton Mullis},
  \bibinfo{person}{Aarush Katta}, \bibinfo{person}{Theo Coombes},
  \bibinfo{person}{Jenia Jitsev}, {and} \bibinfo{person}{Aran Komatsuzaki}.}
  \bibinfo{year}{2021}\natexlab{}.
\newblock \bibinfo{title}{{{LAION-400M}}: {{Open Dataset}} of {{CLIP-Filtered}}
  400 {{Million Image-Text Pairs}}}.
\newblock
\newblock
\urldef\tempurl%
\url{https://doi.org/10.48550/arXiv.2111.02114}
\showDOI{\tempurl}
\showeprint[arxiv]{2111.02114}~[cs]


\bibitem[Scott and Kayali(2020)]%
        {ScottKayali2020}
\bibfield{author}{\bibinfo{person}{Mark Scott} {and} \bibinfo{person}{Laura
  Kayali}.} \bibinfo{year}{2020}\natexlab{}.
\newblock \showarticletitle{What {{Happened When Humans Stopped Managing Social
  Media Content}}}.
\newblock \bibinfo{journal}{\emph{Politico}} (\bibinfo{date}{Oct.}
  \bibinfo{year}{2020}).
\newblock
\urldef\tempurl%
\url{https://www.politico.eu/article/facebook-content-moderation-automation/}
\showURL{%
\tempurl}


\bibitem[Sherman(2021)]%
        {Sherman2021}
\bibfield{author}{\bibinfo{person}{Alex Sherman}.}
  \bibinfo{year}{2021}\natexlab{}.
\newblock \showarticletitle{Reddit's {{CEO}} Has a Colorful Nickname for the
  {{Redditors}} Who Ruin It for Everyone}.
\newblock \bibinfo{journal}{\emph{CNBC}} (\bibinfo{date}{Jan.}
  \bibinfo{year}{2021}).
\newblock
\urldef\tempurl%
\url{https://www.cnbc.com/2021/01/29/reddit-ceo-has-colorful-nickname-for-troublemakers.html}
\showURL{%
\tempurl}


\bibitem[Simon(2007)]%
        {Simon2007}
\bibfield{author}{\bibinfo{person}{Jonathan Simon}.}
  \bibinfo{year}{2007}\natexlab{}.
\newblock \showarticletitle{Rise of the {{Carceral State}}}.
\newblock \bibinfo{journal}{\emph{Social Research}} \bibinfo{volume}{74},
  \bibinfo{number}{2} (\bibinfo{year}{2007}), \bibinfo{pages}{471--508}.
\newblock
\showISSN{0037-783X}
\showeprint[jstor]{40971941}
\urldef\tempurl%
\url{https://www.jstor.org/stable/40971941}
\showURL{%
\tempurl}


\bibitem[Smith~IV(2016)]%
        {SmithIV2016}
\bibfield{author}{\bibinfo{person}{Jack Smith~IV}.}
  \bibinfo{year}{2016}\natexlab{}.
\newblock \bibinfo{title}{China {{Is Creating}} a {{Thoughtcrime-Predicting
  Surveillance System}} for {{Its Citizens}}}.
\newblock
\newblock
\urldef\tempurl%
\url{https://www.mic.com/articles/137464/china-is-creating-a-thoughtcrime-predicting-surveillance-system-for-its-citizens}
\showURL{%
\tempurl}


\bibitem[Spierenburg(2009)]%
        {Spierenburg2009}
\bibfield{author}{\bibinfo{person}{Pieter Spierenburg}.}
  \bibinfo{year}{2009}\natexlab{}.
\newblock \showarticletitle{Stephen {{A}}. {{Toth}}, {{Beyond Papillon}}. {{The
  French}} Overseas Penal Colonies, 1854-1952}.
\newblock \bibinfo{journal}{\emph{Crime, Histoire \& Soci\'et\'es / Crime,
  History \& Societies}} \bibinfo{volume}{13}, \bibinfo{number}{1}
  (\bibinfo{date}{March} \bibinfo{year}{2009}), \bibinfo{pages}{153--155}.
\newblock
\showISBNx{9782600012959}
\showISSN{1422-0857}
\urldef\tempurl%
\url{https://journals.openedition.org/chs/716}
\showURL{%
\tempurl}


\bibitem[Sprick(2019)]%
        {Sprick2019}
\bibfield{author}{\bibinfo{person}{Daniel Sprick}.}
  \bibinfo{year}{2019}\natexlab{}.
\newblock \showarticletitle{Predictive {{Policing}} in {{China}}: {{An
  Authoritarian Dream}} of {{Public Security}}}.
\newblock \bibinfo{journal}{\emph{NAVEI\~N REET: Nordic Journal of Law \&
  Social Research}} \bibinfo{volume}{1}, \bibinfo{number}{9}
  (\bibinfo{year}{2019}), \bibinfo{pages}{299--324}.
\newblock
\showISSN{2246-7807}
\urldef\tempurl%
\url{https://tidsskrift.dk/nnjlsr/article/view/122164/169413}
\showURL{%
\tempurl}


\bibitem[{Stability AI}(2022)]%
        {StabilityAI2022}
\bibfield{author}{\bibinfo{person}{{Stability AI}}.}
  \bibinfo{year}{2022}\natexlab{}.
\newblock \bibinfo{title}{Stable {{Diffusion Public Release}}}.
\newblock
\newblock
\urldef\tempurl%
\url{https://stability.ai/blog/stable-diffusion-public-release}
\showURL{%
\tempurl}


\bibitem[Stjernfelt and Lauritzen(2019)]%
        {StjernfeltLauritzen2019}
\bibfield{author}{\bibinfo{person}{Frederik Stjernfelt} {and}
  \bibinfo{person}{Anne~Mette Lauritzen}.} \bibinfo{year}{2019}\natexlab{}.
\newblock \bibinfo{booktitle}{\emph{Your {{Post}} Has Been {{Removed}}: {{Tech
  Giants}} and {{Freedom}} of {{Speech}}}}.
\newblock \bibinfo{publisher}{Springer International Publishing},
  \bibinfo{address}{Cham}.
\newblock
\showISBNx{978-3-030-25968-6}
\urldef\tempurl%
\url{https://doi.org/10.1007/978-3-030-25968-6}
\showDOI{\tempurl}


\bibitem[Stockton(2020)]%
        {Stockton2020}
\bibfield{author}{\bibinfo{person}{Nick Stockton}.}
  \bibinfo{year}{2020}\natexlab{}.
\newblock \showarticletitle{Incident 78: {{Meet}} the {{Secret Algorithm
  That}}'s {{Keeping Students Out}} of {{College}}}.
\newblock \bibinfo{journal}{\emph{Artificial Intelligence Incident Database}}
  (\bibinfo{date}{July} \bibinfo{year}{2020}).
\newblock
\urldef\tempurl%
\url{https://incidentdatabase.ai/cite/78}
\showURL{%
\tempurl}


\bibitem[{Students at the University of Bristol}(2021)]%
        {StudentsAtTheUniversityOfBristol2021}
\bibfield{author}{\bibinfo{person}{{Students at the University of Bristol}}.}
  \bibinfo{year}{2021}\natexlab{}.
\newblock \bibinfo{title}{The {{Dreyfus Affair}} and the {{Image}} of the
  {{Intellectual}}}.
\newblock
\newblock
\urldef\tempurl%
\url{https://intellectualsandthemedia.org/2021/12/11/the-dreyfus-affair-and-the-image-of-the-intellectual/}
\showURL{%
\tempurl}


\bibitem[Suebsaeng and Reis(2023)]%
        {SuebsaengReis2023}
\bibfield{author}{\bibinfo{person}{Asawin Suebsaeng} {and}
  \bibinfo{person}{Patrick Reis}.} \bibinfo{year}{2023}\natexlab{}.
\newblock \showarticletitle{Trump's {{Killing Spree}}: {{The Inside Story}} of
  {{His Race}} to {{Execute Every Prisoner He Could}}}.
\newblock \bibinfo{journal}{\emph{Rolling Stone}} (\bibinfo{date}{Jan.}
  \bibinfo{year}{2023}).
\newblock
\urldef\tempurl%
\url{https://www.rollingstone.com/politics/politics-features/trump-capital-punishment-brandon-bernard-lisa-montgomery-1234664126/}
\showURL{%
\tempurl}


\bibitem[Supran and Rahmstorf(2023)]%
        {SupranRahmstorf2023}
\bibfield{author}{\bibinfo{person}{G Supran} {and} \bibinfo{person}{S
  Rahmstorf}.} \bibinfo{year}{2023}\natexlab{}.
\newblock \showarticletitle{Assessing {{ExxonMobil}}'s Global Warming
  Projections}.
\newblock \bibinfo{journal}{\emph{Science}} \bibinfo{volume}{379},
  \bibinfo{number}{6628} (\bibinfo{date}{Jan.} \bibinfo{year}{2023}).
\newblock
\urldef\tempurl%
\url{https://doi.org/10.1126/science.abk0063}
\showDOI{\tempurl}


\bibitem[Tabuchi(2021)]%
        {Tabuchi2021}
\bibfield{author}{\bibinfo{person}{Hiroko Tabuchi}.}
  \bibinfo{year}{2021}\natexlab{}.
\newblock \showarticletitle{In {{Your Facebook Feed}}: {{Oil Industry Pushback
  Against Biden Climate Plans}}}.
\newblock \bibinfo{journal}{\emph{The New York Times}} (\bibinfo{date}{Sept.}
  \bibinfo{year}{2021}).
\newblock
\showISSN{0362-4331}
\urldef\tempurl%
\url{https://www.nytimes.com/2021/09/30/climate/api-exxon-biden-climate-bill.html}
\showURL{%
\tempurl}


\bibitem[Taylor and Watts(2019)]%
        {TaylorWatts2019}
\bibfield{author}{\bibinfo{person}{Matthew Taylor} {and}
  \bibinfo{person}{Jonathan Watts}.} \bibinfo{year}{2019}\natexlab{}.
\newblock \showarticletitle{Revealed: The 20 Firms behind a Third of All Carbon
  Emissions}.
\newblock \bibinfo{journal}{\emph{The Guardian}} (\bibinfo{date}{Oct.}
  \bibinfo{year}{2019}).
\newblock
\showISSN{0261-3077}
\urldef\tempurl%
\url{https://www.theguardian.com/environment/2019/oct/09/revealed-20-firms-third-carbon-emissions}
\showURL{%
\tempurl}


\bibitem[Tolich(2010)]%
        {Tolich2010}
\bibfield{author}{\bibinfo{person}{Martin Tolich}.}
  \bibinfo{year}{2010}\natexlab{}.
\newblock \showarticletitle{A {{Critique}} of {{Current Practice}}: {{Ten
  Foundational Guidelines}} for {{Autoethnographers}}}.
\newblock \bibinfo{journal}{\emph{Qualitative Health Research}}
  \bibinfo{volume}{20}, \bibinfo{number}{12} (\bibinfo{date}{Dec.}
  \bibinfo{year}{2010}), \bibinfo{pages}{1599--1610}.
\newblock
\showISSN{1049-7323, 1552-7557}
\urldef\tempurl%
\url{https://doi.org/10.1177/1049732310376076}
\showDOI{\tempurl}


\bibitem[{Twitter}(2022)]%
        {Twitter2022}
\bibfield{author}{\bibinfo{person}{{Twitter}}.}
  \bibinfo{year}{2022}\natexlab{}.
\newblock \bibinfo{booktitle}{\emph{Rules {{Enforcement}}}}.
\newblock \bibinfo{type}{Transparency {{Report}}}~20.
  \bibinfo{institution}{Twitter}.
\newblock
\urldef\tempurl%
\url{https://transparency.twitter.com/en/reports/rules-enforcement.html}
\showURL{%
\tempurl}


\bibitem[Tyler(2003)]%
        {Tyler2003}
\bibfield{author}{\bibinfo{person}{Tom~R. Tyler}.}
  \bibinfo{year}{2003}\natexlab{}.
\newblock \showarticletitle{Procedural {{Justice}}, {{Legitimacy}}, and the
  {{Effective Rule}} of {{Law}}}.
\newblock \bibinfo{journal}{\emph{Crime and Justice}}  \bibinfo{volume}{30}
  (\bibinfo{year}{2003}), \bibinfo{pages}{283--357}.
\newblock
\showISSN{0192-3234}
\showeprint[jstor]{1147701}
\urldef\tempurl%
\url{https://www.jstor.org/stable/1147701}
\showURL{%
\tempurl}


\bibitem[Tyler(2008)]%
        {Tyler2007}
\bibfield{author}{\bibinfo{person}{Tom~R Tyler}.}
  \bibinfo{year}{2007--2008}\natexlab{}.
\newblock \showarticletitle{Procedural {{Justice}} and the {{Courts}}}.
\newblock \bibinfo{journal}{\emph{Court Review}} \bibinfo{volume}{44},
  \bibinfo{number}{1--2} (\bibinfo{year}{2007--2008}), \bibinfo{pages}{26--31}.
\newblock
\urldef\tempurl%
\url{https://amjudges.org/publications/courtrv/cr44-1/CR44-1-2Tyler.pdf}
\showURL{%
\tempurl}


\bibitem[Tyler(1200)]%
        {Tyler2006}
\bibfield{author}{\bibinfo{person}{Tom~R. Tyler}.} \bibinfo{year}{Sun,
  05/07/2006 - 12:00}\natexlab{}.
\newblock \bibinfo{booktitle}{\emph{Why {{People Obey}} the {{Law}}}}.
\newblock \bibinfo{publisher}{Princeton University Press},
  \bibinfo{address}{Princeton, NJ, USA}.
\newblock
\showISBNx{978-0-691-12673-9}
\urldef\tempurl%
\url{https://press.princeton.edu/books/paperback/9780691126739/why-people-obey-the-law}
\showURL{%
\tempurl}


\bibitem[{Vann R Newkirk II}(2025)]%
        {VannRNewkirkII2025}
\bibfield{author}{\bibinfo{person}{{Vann R Newkirk II}}.}
  \bibinfo{year}{2025}\natexlab{}.
\newblock \bibinfo{title}{What {{Climate Change Will Do}} to {{America}} by
  {{Mid-Century}}}.
\newblock
\newblock
\urldef\tempurl%
\url{https://www.theatlantic.com/magazine/2025/12/trump-climate-change-acceleration/684632/}
\showURL{%
\tempurl}


\bibitem[Vasist et~al\mbox{.}(2024)]%
        {VasistChatterjeeea2024}
\bibfield{author}{\bibinfo{person}{Pramukh~Nanjundaswamy Vasist},
  \bibinfo{person}{Debashis Chatterjee}, {and} \bibinfo{person}{Satish
  Krishnan}.} \bibinfo{year}{2024}\natexlab{}.
\newblock \showarticletitle{The {{Polarizing Impact}} of {{Political
  Disinformation}} and {{Hate Speech}}: {{A Cross-country Configural
  Narrative}}}.
\newblock \bibinfo{journal}{\emph{Information Systems Frontiers}}
  \bibinfo{volume}{26}, \bibinfo{number}{2} (\bibinfo{date}{April}
  \bibinfo{year}{2024}), \bibinfo{pages}{663--688}.
\newblock
\showISSN{1387-3326, 1572-9419}
\urldef\tempurl%
\url{https://doi.org/10.1007/s10796-023-10390-w}
\showDOI{\tempurl}


\bibitem[Wallechinsky and Wallace(1978)]%
        {WallechinskyWallace1978}
\bibfield{author}{\bibinfo{person}{David Wallechinsky} {and}
  \bibinfo{person}{Irving Wallace}.} \bibinfo{year}{1978}\natexlab{}.
\newblock \bibinfo{title}{Famous {{Lasts}}: {{The Last Prisoner}} on
  {{Devil}}'s {{Island}}}.
\newblock
\newblock
\showISBNx{0-688-03372-5}
\urldef\tempurl%
\url{https://archive.org/details/peoplesalmanac200wall}
\showURL{%
\tempurl}


\bibitem[{Wikipedia}(2023)]%
        {Wikipedia2023}
\bibfield{editor}{\bibinfo{person}{{Wikipedia}}} (Ed.).
  \bibinfo{year}{2023}\natexlab{}.
\newblock \showarticletitle{Francis {{Bacon}} (Artist)}.
\newblock \bibinfo{journal}{\emph{Wikipedia}} (\bibinfo{date}{Jan.}
  \bibinfo{year}{2023}).
\newblock
\urldef\tempurl%
\url{https://en.wikipedia.org/wiki/Francis_Bacon_(artist)}
\showURL{%
\tempurl}


\bibitem[{Wikipedia}(2025)]%
        {WikipediaTechFirmsByRevenue2025}
\bibfield{editor}{\bibinfo{person}{{Wikipedia}}} (Ed.).
  \bibinfo{year}{2025}\natexlab{}.
\newblock \showarticletitle{List of Largest Technology Companies by Revenue}.
\newblock \bibinfo{journal}{\emph{Wikipedia}} (\bibinfo{date}{Dec.}
  \bibinfo{year}{2025}).
\newblock
\urldef\tempurl%
\url{https://en.wikipedia.org/w/index.php?title=List_of_largest_technology_companies_by_revenue&oldid=1329579749}
\showURL{%
\tempurl}


\bibitem[{Wikipedia}(2026)]%
        {WikipediaCountriesByGDP2026}
\bibfield{editor}{\bibinfo{person}{{Wikipedia}}} (Ed.).
  \bibinfo{year}{2026}\natexlab{}.
\newblock \showarticletitle{List of Countries by {{GDP}} (Nominal)}.
\newblock \bibinfo{journal}{\emph{Wikipedia}} (\bibinfo{date}{Jan.}
  \bibinfo{year}{2026}).
\newblock
\urldef\tempurl%
\url{https://en.wikipedia.org/w/index.php?title=List_of_countries_by_GDP_(nominal)&oldid=1331072395}
\showURL{%
\tempurl}


\bibitem[{Worldometer}(2023)]%
        {Worldometer2023}
\bibfield{author}{\bibinfo{person}{{Worldometer}}.}
  \bibinfo{year}{2023}\natexlab{}.
\newblock \bibinfo{title}{United {{States Population}}}.
\newblock
\newblock
\urldef\tempurl%
\url{https://www.worldometers.info/world-population/us-population/}
\showURL{%
\tempurl}


\bibitem[Yampolskiy(2015)]%
        {Yampolskiy2015}
\bibfield{author}{\bibinfo{person}{Roman Yampolskiy}.}
  \bibinfo{year}{2015}\natexlab{}.
\newblock \showarticletitle{Incident 57: {{Australian Automated Debt Assessment
  System Issued False Notices}} to {{Thousands}}}.
\newblock \bibinfo{journal}{\emph{Artificial Intelligence Incident Database}}
  (\bibinfo{date}{July} \bibinfo{year}{2015}).
\newblock
\urldef\tempurl%
\url{https://incidentdatabase.ai/cite/57/}
\showURL{%
\tempurl}


\bibitem[Yampolskiy(2016)]%
        {Yampolskiy2016}
\bibfield{author}{\bibinfo{person}{Roman Yampolskiy}.}
  \bibinfo{year}{2016}\natexlab{}.
\newblock \showarticletitle{Incident 11: {{Northpointe Risk Models}}}.
\newblock \bibinfo{journal}{\emph{Artificial Intelligence Incident Database}}
  (\bibinfo{date}{May} \bibinfo{year}{2016}).
\newblock
\urldef\tempurl%
\url{https://incidentdatabase.ai/cite/11}
\showURL{%
\tempurl}


\bibitem[Yang(2021)]%
        {Yang2021}
\bibfield{author}{\bibinfo{person}{Jisheng Yang}.}
  \bibinfo{year}{2021}\natexlab{}.
\newblock \bibinfo{booktitle}{\emph{The {{World Turned Upside Down}}: {{A
  History}} of the {{Chinese Cultural Revolution}}}}.
\newblock \bibinfo{publisher}{{Farrar, Straus and Giroux}},
  \bibinfo{address}{New York, NY, USA}.
\newblock
\showISBNx{978-0-374-29313-0}
\urldef\tempurl%
\url{https://us.macmillan.com/books/9781250829702/theworldturnedupsidedown}
\showURL{%
\tempurl}


\end{thebibliography}

\appendix

\clearpage
\section{Policy and Screenshot for Twitter}
\label{app:twitter}

\REF{app:twitter-abusive-behavior} documents Twitter's policy on abusive
behavior
\href{https://web.archive.org/web/20211022141522/https://help.twitter.com/en/rules-and-policies/abusive-behavior}{as
of 22 October, 2021}. It preserves the structure of the original text, with
links pointing to the corresponding pages in the Internet Archive.
\REF{app:twitter-staging} documents Twitter's user interface for a locked
account with the violative tweet at the center.

\subsection{Twitter's Policy on Abusive Behavior}
\label{app:twitter-abusive-behavior}

\noindent\href{https://web.archive.org/web/20211022141522/https://help.twitter.com/content/help-twitter/en/rules-and-policies/twitter-rules}{Twitter
Rules}: You may not engage in the targeted harassment of someone, or incite
other people to do so. We consider abusive behavior an attempt to harass,
intimidate, or silence someone else's voice.

\subsubsection{Rationale}

On Twitter, you should feel safe expressing your unique point of view. We
believe in freedom of expression and open dialogue, but that means little as an
underlying philosophy if voices are silenced because people are afraid to speak
up.

In order to facilitate healthy dialogue on the platform, and empower individuals
to express diverse opinions and beliefs, we prohibit behavior that harasses or
intimidates, or is otherwise intended to shame or degrade others. In addition to
posing risks to people's safety, abusive behavior may also lead to physical and
emotional hardship for those affected.

Learn more about our approach to
\href{https://web.archive.org/web/20211022141522/https://help.twitter.com/content/help-twitter/en/rules-and-policies/enforcement-philosophy}{policy
development and our enforcement philosophy}.

\subsubsection{When This Applies}

Some Tweets may seem to be abusive when viewed in isolation, but may not be when
viewed in the context of a larger conversation. When we review this type of
content, it may not be clear whether it is intended to harass an individual, or
if it is part of a consensual conversation. To help our teams understand the
context of a conversation, we may need to hear directly from the person being
targeted, to ensure that we have the information needed prior to taking any
enforcement action.

We will review and take action against reports of accounts targeting an
individual or group of people with any of the following behavior within Tweets
or Direct Messages. For accounts engaging in abusive behavior on their profile,
please refer to our
\href{https://web.archive.org/web/20211022141522/https://help.twitter.com/content/help-twitter/en/rules-and-policies/abusive-profile}{abusive
profile policy}. For behavior targeting people based on their race, ethnicity,
national origin, sexual orientation, gender, gender identity, religious
affiliation, age, disability, or serious disease, this may be in violation of
our
\href{https://web.archive.org/web/20211022141522/https://help.twitter.com/content/help-twitter/en/rules-and-policies/hateful-conduct-policy}{hateful
conduct policy}.

\begin{description}

\item[Violent Threats] \hfill

    We prohibit content that makes violent threats against an identifiable
    target. Violent threats are declarative statements of intent to inflict
    injuries that would result in serious and lasting bodily harm, where an
    individual could die or be significantly injured, e.g., ``I will kill you.''

    \textbf{Note}: We have a zero tolerance policy against violent threats.
    Those deemed to be sharing violent threats will face immediate and permanent
    suspension of their account.

\item[Wishing, hoping, or calling for serious harm on a person or group of
    people] \hfill

    We do not tolerate content that wishes, hopes, promotes, incites, or
    expresses a desire for death, serious bodily harm or serious disease against
    an individual or group of people. This includes, but is not limited to:

    \begin{itemize}
    \item Hoping that someone dies as a result of a serious disease e.g., ``I
        hope you get cancer and die.''
    \item Wishing for someone to fall victim to a serious accident e.g., ``I
        wish that you would get run over by a car next time you run your
        mouth.''
    \item Saying that a group of individuals deserves serious physical injury
        e.g., ``If this group of protesters don't shut up, they deserve to be
        shot.''
    \end{itemize}

\item[About wishes of harm exceptions on Twitter] \hfill

    We recognize that conversations regarding certain individuals credibly
    accused of severe violence may prompt outrage and associated wishes of harm.
    In these limited cases, we will request the user to delete the Tweet without
    any risk of account penalty, strike, or suspension. Examples are, but not
    limited to:

    \begin{itemize}
    \item ``I wish all rapists to die.''
    \item ``Child abusers should be hanged.''
    \end{itemize}

\item[Unwanted sexual advances] \hfill

    While some
    \href{https://web.archive.org/web/20211022141522/https://help.twitter.com/content/help-twitter/en/rules-and-policies/media-policy}{consensual
    nudity and adult content is permitted} on Twitter, we prohibit unwanted
    sexual advances and content that sexually objectifies an individual without
    their consent. This includes, but is not limited to:

    \begin{itemize}
    \item sending someone unsolicited and/or unwanted adult media, including
        images, videos, and GIFs;
    \item unwanted sexual discussion of someone's body;
    \item solicitation of sexual acts; and
    \item any other content that otherwise sexualizes an individual without
        their consent.
    \end{itemize}

\item[Using insults, profanity, or slurs with the purpose of harassing or
    intimidating others] \hfill

    We take action against the use of insults, profanity, or slurs to target
    others. In some cases, such as (but not limited to) severe, repetitive usage
    of insults or slurs where the primary intent is to harass or intimidate
    others, we may require Tweet removal. In other cases, such as (but not
    limited to) moderate, isolated usage of insults and profanity where the
    primary intent is to harass or intimidate others, we may limit Tweet
    visibility as further described below. Please also note that while some
    individuals may find certain terms to be offensive, we will not take action
    against every instance where insulting terms are used.

\item[Encouraging or calling for others to harass an individual or group of
    people] \hfill

    We prohibit behavior that encourages others to harass or target specific
    individuals or groups with abusive behavior. This includes, but is not
    limited to; calls to target people with abuse or harassment online and
    behavior that urges offline action such as physical harassment.

\item[Denying mass casualty events took place] \hfill

    We prohibit content that denies that mass murder or other mass casualty
    events took place, where we can verify that the event occured [sic], and

    when the content is shared with abusive intent. This may include references
    to such an event as a ``hoax'' or claims that victims or survivors are fake
    or ``actors.'' It includes, but is not limited to, events like the
    Holocaust, school shootings, terrorist attacks, and natural disasters.

\item[Do I need to be the target of this content for it to be reviewed for
    violating the Twitter Rules?] \hfill

    No, we review both first-person and bystander reports of such content.
\end{description}

\subsubsection{Consequences}

When determining the penalty for violating this policy, we consider a number of
factors including, but not limited to, the severity of the violation and an
individual's previous record of rule violations. The following is a list of
potential enforcement options for content that violates this policy:
\begin{itemize}
\item Downranking Tweets in replies, except when the user follows the Tweet
    author.
\item Making Tweets ineligible for amplification in Top search results and/or on
    timelines for users who don't follow the Tweet author.
\item Excluding Tweets and/or accounts in email or in-product recommendations.
\item Requiring Tweet removal.
    \begin{itemize}
    \item For example, we may ask someone to remove the violating content and
        serve a period of time in read-only mode before they can Tweet again.
        Subsequent violations will lead to longer read- only periods and may
        eventually result in permanent suspension.
    \end{itemize}
\item Suspending accounts whose primary use we've determined is to engage in
    abusive behavior as defined in this policy, or who have shared violent
    threats.
\end{itemize}
Learn more about
\href{https://web.archive.org/web/20211022141522/https://help.twitter.com/content/help-twitter/en/rules-and-policies/enforcement-options}{our
range of enforcement options}.

If someone believes their account was suspended in error, they can
\href{https://web.archive.org/web/20211022141522/https://help.twitter.com/forms/general?subtopic=suspended}{submit
an appeal}.

\subsection{Twitter's Restricted UI}
\label{app:twitter-staging}

\begin{figure}
\centering
\includegraphics[width=0.6\textwidth]{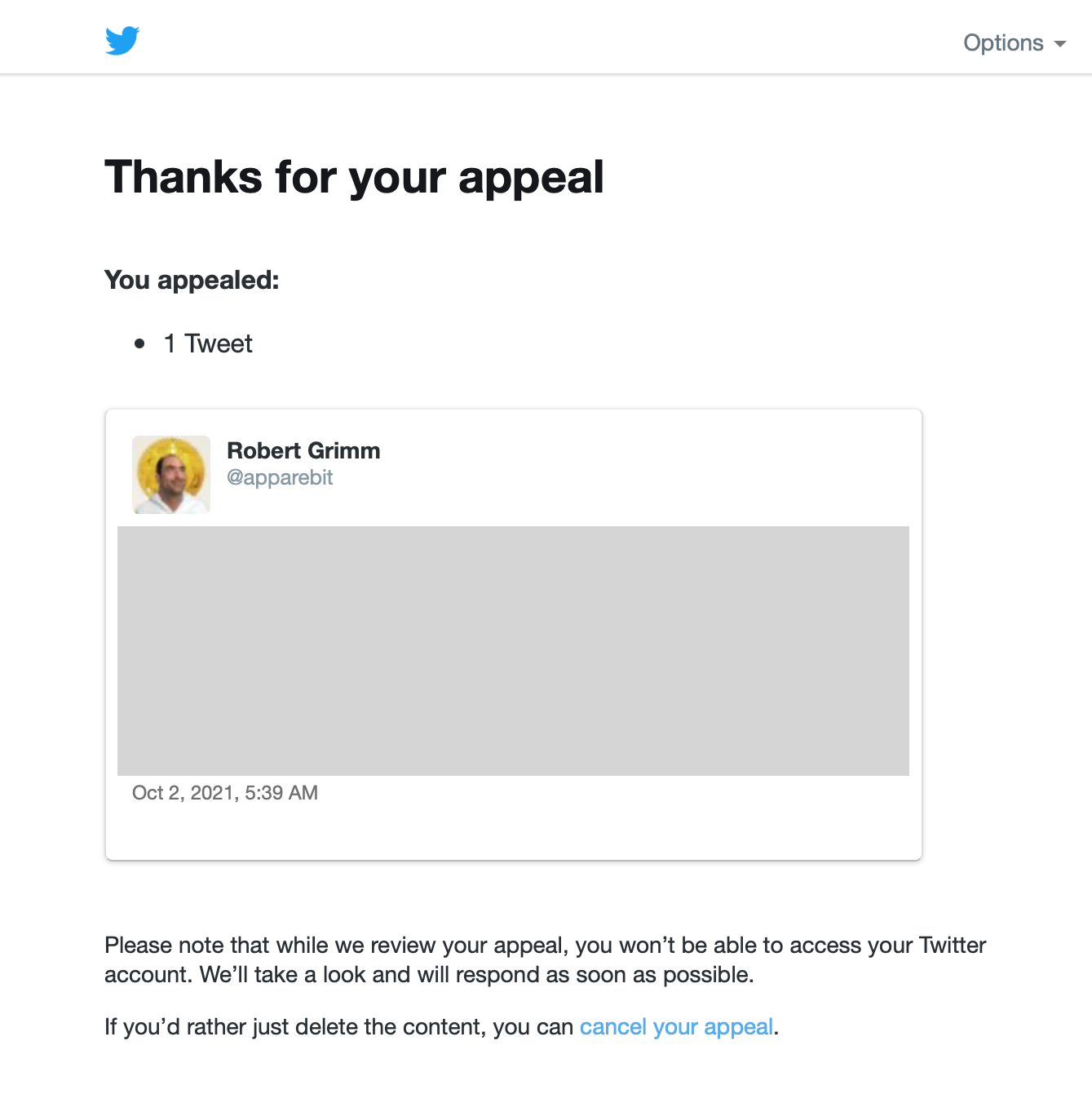}
\Description{Screenshot showing the violative tweet in the center under the
    headline ``Thanks for your appeal.''}
\caption{Twitter's restricted user interface centers all attention on the
violative tweet.}
\label{fig:thetweet}
\end{figure}

Figure~\ref{fig:thetweet} illustrates Twitter's browser-based user interface
when an account is locked down. The tweet's actual text is elided for its
incivility; its substance is described at the start of
\REF{sec:tweet-da-fe}. The violative tweet centers all attention. It remains
in place even after cancelling the appeal but before agreeing to delete the
tweet; only the text above and below changes. Twitter's original instructions
claimed that ``while in this state, you can still browse Twitter, but you're
limited to only sending Direct Messages to your followers---no Tweets, Retweets,
Fleets, follows, or likes.'' In October 2021, none of that was possible while an
appeal is pending.
 \clearpage
\section{Policy and Terms of Use for \DDAALLEE\ 2}
\label{app:dalle}

\begin{figure}
\centering
\includegraphics[width=0.3\textwidth]{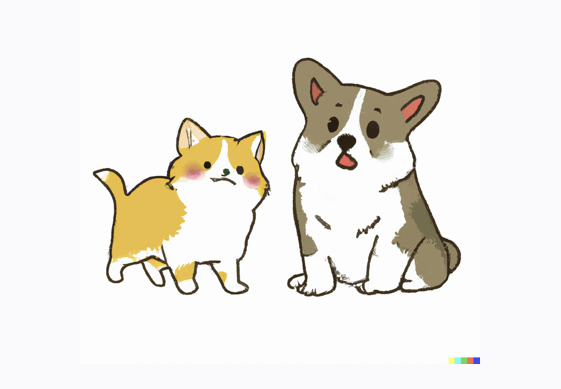}
\Description{Cartoon of sad kitten and puppy}
\caption{Be nice to \DALLE's pets! \copyright\ OpenAI}\label{fig:dalle-cartoon}
\end{figure}

\REF{app:dalle-content-policy} documents \DALLE~2's content policy and
\REF{app:dalle:terms} its addendum to OpenAI's terms of use, both as of July
20, 2022. They preserve the structure of the original, with links pointing to
pages in the Internet Archive. Note that archived OpenAI webpages may contain
\V{CSS} that prevents the printing of a full page and JavaScript that redirects
to an error page after a few seconds.

The content policy was located at
\url{https://labs.openai.com/policies/content-policy}. It was updated on
September 19, 2022 by rewording the rules on disclosing the role of \V{AI} and
by removing the fourth bullet of the rules on respecting the rights of others.
At that time, OpenAI also updated its notification for violative prompts to
state ``It looks like this request may not follow our content policy.'' above
the cartoon shown in Fig.~\ref{fig:dalle-cartoon}, a rather abrupt switch from
the inappropriately punitive to the inappropriately saccharine.

The addendum to OpenAI's terms of use was located at
\url{https://labs.openai.com/policies/terms}, but was rescinded on November 4,
2022.

\subsection{\DDAALLEE\ Content Policy}
\label{app:dalle-content-policy}

Thank you for trying our generative \V{AI} tools!

\noindent In your usage, you must adhere to our Content Policy:

\begin{description}
\item[Do not attempt to create, upload, or share images that are not G-rated or
    that could cause harm.] \hfill

    \begin{itemize}
    \item \textbf{Hate}: hateful symbols, negative stereotypes, comparing certain
        groups to animals/objects, or otherwise expressing or promoting hate based
        on identity.
    \item \textbf{Harassment}: mocking, threatening, or bullying an individual.
    \item \textbf{Violence}: violent acts and the suffering or humiliation of
        others.
    \item \textbf{Self-harm}: suicide, cutting, eating disorders, and other attempts
        at harming oneself.
    \item \textbf{Sexual}: nudity, sexual acts, sexual services, or content
        otherwise meant to arouse sexual excitement.
    \item \textbf{Shocking}: bodily fluids, obscene gestures, or other profane
        subjects that may shock or disgust.
    \item \textbf{Illegal activity}: drug use, theft, vandalism, and other illegal
        activities.
    \item \textbf{Deception}: major conspiracies or events related to major ongoing
        geopolitical events.
    \item \textbf{Political}: politicians, ballot-boxes, protests, or other content
        that may be used to influence the political process or to campaign.
    \item \textbf{Public and personal health}: the treatment, prevention, diagnosis,
        or transmission of diseases, or people experiencing health ailments.
    \item \textbf{Spam}: unsolicited bulk content.
    \end{itemize}

\item[Disclose the role of \V{AI}.] \hfill

    \begin{itemize}
    \item You must clearly indicate that images are \V{AI}-generated---or which
        portions of them are---by attributing to OpenAI when sharing, whether in
        public or private.
    \item You may post these images to social media. Please refer to our
        \href{https://web.archive.org/web/20220803232350/https://openai.com/api/policies/sharing-publication/}{Sharing
        and Publication Policy} for further details.
    \end{itemize}

\item[Respect the rights of others.] \hfill

    \begin{itemize}
    \item Do not upload images of people without their consent, including public
        figures.
    \item Do not upload images to which you do not hold appropriate usage rights.
    \item Do not attempt to create images of public figures (including celebrities).
    \item To prevent deepfakes, we are currently prohibiting uploads of all
        realistic faces, even when the face belongs to you or if you have consent.
    \end{itemize}

\item[Please report any suspected violations of these rules to our Support team
    (\url{support@openai.com}).] \hfill

    \begin{itemize}
    \item We will investigate and take action accordingly, up to and including
        terminating the violating account.
    \end{itemize}

\end{description}

\subsection{\DDAALLEE\ Terms of Use}
\label{app:dalle:terms}

Thank you for your interest in \DALLE. Access to \DALLE\ is subject to OpenAI's
\href{https://web.archive.org/web/20220729134013/https://openai.com/api/policies/terms/}{Terms
of Use} and the additional terms below. By using \DALLE, you agree to these
terms.

\begin{EnumerateLining}
\item \textbf{Use of \DALLE.} \DALLE\ can generate images (``Generations'')
    based on text input you provide (``Prompts''). You may also upload images to
    \DALLE\ (``Uploads'') and create Generations with Uploads.
\item \textbf{Use of Images.} Subject to your compliance with these terms and
    our Content Policy, you may use Generations for any legal purpose, including
    for commercial use. This means you may sell your rights to the Generations
    you create, incorporate them into works such as books, websites, and
    presentations, and otherwise commercialize them.
\item \textbf{Buying Credits.} You may buy credits to create additional
    Generations, subject to the payment terms in our Terms of Use. Credits must
    be used within one year of purchase or they will expire. We may change our
    prices at any time by updating our pricing page.
\item \textbf{No Infringing or Harmful Use.} You must comply with our Content
    Policy, and you may not use \DALLE\ in a way that may harm a person or
    infringe their rights. For example, you may not submit Uploads for which you
    don't have the necessary rights, images of people without their consent, or
    Prompts intended to generate harmful or illegal images. We may delete
    Prompts and Uploads, or suspend or ban your account for any violations. You
    may not seek to reverse engineer \DALLE, use \DALLE\ to attempt to build a
    competitive product or service, or otherwise infringe our rights. You will
    indemnify us for your use of \DALLE\ as outlined in our Terms of Use.
\item \textbf{Improving \V{AI} safety and technologies.} You grant us all rights
    to use your Prompts and Uploads to improve our \V{AI} safety efforts, and to
    develop and improve our \V{AI} technologies, products, and services. As part
    of this, Prompts and Uploads may be shared with and manually reviewed by a
    person (for example, if a Generation is flagged as sensitive), including by
    third party contractors located around the world. You should not provide any
    Prompts or Uploads that are sensitive or that you do not want others to
    view, including Prompts or Uploads that include personal data. You can
    request deletion of Uploads by contacting \url{support@openai.com}.
\item \textbf{Ownership of Generations.} To the extent allowed by law and as
    between you and OpenAI, you own your Prompts and Uploads, and you agree that
    OpenAI owns all Generations (including Generations with Uploads but not the
    Uploads themselves), and you hereby make any necessary assignments for this.
    OpenAI grants you the exclusive rights to reproduce and display such
    Generations and will not resell Generations that you have created, or assert
    any copyright in such Generations against you or your end users, all
    provided that you comply with these terms and our Content Policy. If you
    violate our terms or Content Policy, you will lose rights to use
    Generations, but we will provide you written notice and a reasonable
    opportunity to fix your violation, unless it was clearly illegal or abusive.
    You understand and acknowledge that similar or identical Generations may be
    created by other people using their own Prompts, and your rights are only to
    the specific Generation that you have created.
\item \textbf{No Guarantees.} We plan to continue to develop and improve \DALLE,
    but we make no guarantees or promises about how \DALLE\ operates or that it
    will function as intended, and your use of \DALLE is at your own risk.
    Contact \url{support@openai.com} with any questions about your account, or
    \url{dalle-policy@openai.com} with general questions or feedback about use
    of the technology.
\end{EnumerateLining}
 \clearpage
\section{Notification Email and Policy for Pinterest}
\label{app:pinterest}

\REF{app:pinterest:notification} documents Pinterest's email notifying about
the take-down of adult content in early March 2024, and
\REF{app:pinterest:policy} documents the corresponding section from the
Pinterest's community guidelines. Both preserve the structure of the original.
Public links in the email point to pages in the Internet Archive, whereas
private links point to \texttt{about:blank}.

\subsection{Pinterest's Notification Email}
\label{app:pinterest:notification}

\textbf{Hi\ ,} [sic]

We recently removed a Pin from your board "[elided for privacy]" for violation
of our
\href{https://web.archive.org/web/20240307064016/https://policy.pinterest.com/en/community-guidelines#section-adult-content}{Community
Guidelines} on adult content.

To see the Pin we removed, use this \href{about:blank}{one-time link} (please
note, the link will expire after 7 days). And if you think we've made a mistake,
you can \href{about:blank}{submit an appeal} within 7 days.

We remove or limit the distribution of mature and explicit content, including:

\begin{itemize}
    \item Nudity
    \item Sexualized content, even if the people are clothed or partially
    clothed
    \item Graphic depictions of sexual activity in imagery or text
    \item Fetish imagery
\end{itemize}

These rules apply to all Pins, including ones on your secret boards. Please take
some time to go through your Pins and remove any that may be in violation of our
\href{https://web.archive.org/web/20240307064016/https://policy.pinterest.com/en/community-guidelines#section-adult-content}{Community
Guidelines}, or we may take additional action on your account.

If you think we've made a mistake, you can \href{about:blank}{submit an appeal}
within 7 days.

Thanks,

\textbf{The Pinterest Team}

\subsection{Pinterest's Community Guidelines on Adult Content}
\label{app:pinterest:policy}

Pinterest isn't a place for adult content, including pornography and most
nudity. We remove or limit the distribution of mature and explicit content,
including:

\begin{itemize}
    \item Nudity
    \item Sexualized content, even if the people are clothed or partially
        clothed
    \item Graphic depictions of sexual activity in imagery or text
    \item Fetish imagery
\end{itemize}

We allow some nudity in certain contexts, although we may limit its
distribution. For instance, nudity in paintings and sculptures and in science
and historical contexts is okay. Content about breastfeeding and mastectomies is
also allowed. These guidelines apply to text and real-life, digital, and
animated images and video.
 
\end{document}